\documentclass[12pt]{article}
\usepackage[margin=1in]{geometry}
\usepackage{amssymb, amsmath, amsthm} 
\usepackage{enumerate}
\usepackage{graphicx}
\usepackage{authblk}

\usepackage{setspace}
\newcommand{\indep}{\rotatebox[origin=c]{90}{$\models$}}

\RequirePackage{natbib}
\RequirePackage[colorlinks]{hyperref}
\RequirePackage{hypernat}

\theoremstyle{definition}

\theoremstyle{definition}

\theoremstyle{remark}

\theoremstyle{plain}
\newtheorem{assump}{Assumption}[section]

\newtheorem{lem}{Lemma}[section]
\newtheorem{thm}{Theorem}[section]
\newtheorem{cor}{Corollary}[section]

\makeatletter \renewenvironment{proof}[1][\proofname] {\par\pushQED{\qed}\normalfont\topsep6\p@\@plus6\p@\relax\trivlist\item[\hskip\labelsep\bfseries#1\@addpunct{.}]\ignorespaces}{\popQED\endtrivlist\@endpefalse} \makeatother

\title{Identifying the Effect of a Mis-classified, Binary, Endogenous Regressor\thanks{We thank Daron Acemoglu, Manuel Arellano, Kristy Buzard, Xu Cheng, Bernardo da Silveira, Bo Honor\'{e}, Arthur Lewbel, Chuck Manski, Sophocles Mavroeidis, Francesca Molinari, Yuya Takahashi, the associate editor, two anonymous referees, and seminar participants at Cambridge, CEMFI, Chicago Booth, Manchester, Northwestern, Oxford, Penn State, Princeton, UCL, the 2016 Greater New York Area Econometrics Colloquium, Camp Econometrics IX, and the 2017 North American Summer Meeting of the Econometric Society for valuable comments and suggestions. This document supersedes an earlier version entitled ``On Mis-measured Binary Regressors: New Results and Some Comments on the Literature.''}}

\author[1]{Francis J.\ DiTraglia and Camilo Garc\'{i}a-Jimeno}

\date{\small Final Version: January 23, 2019} 
\begin{document}

\clearpage
\maketitle
\thispagestyle{empty}

\begin{abstract}
  \singlespacing
This paper studies identification of the effect of a mis-classified, binary, endogenous regressor when a discrete-valued instrumental variable is available. 
We begin by showing that the only existing point identification result for this model is incorrect. 
We go on to derive the sharp identified set under mean independence assumptions for the instrument and measurement error.
The resulting bounds are novel and informative, but fail to point identify the effect of interest.
This motivates us to consider alternative and slightly stronger assumptions: we show that adding second and third moment independence assumptions suffices to identify the model.


\end{abstract}

\newpage
\setcounter{page}{1}

\singlespacing
\section{Introduction}

Measurement error and endogeneity are pervasive features of economic data.
Conveniently, a valid instrumental variable corrects for both problems when the measurement error is classical, i.e.\ uncorrelated with the true value of the regressor.
Many regressors of interest in applied work, however, are binary and thus cannot be subject to classical measurement error.\footnote{The only way to mis-classify a true one is downwards, as a zero, while the only way to mis-classify a true zero is upwards, as a one. This creates negative dependence between the truth and measurement error.}
When faced with non-classical measurement error, the instrumental variables estimator can be severely biased.
In this paper, we study an additively separable model of the form
\begin{equation}
  y = c(\mathbf{x}) + \beta(\mathbf{x}) T^* + \varepsilon
  \label{eq:model}
\end{equation}
where $\varepsilon$ is a mean-zero error term, $T^*$ is a binary, potentially endogenous regressor of interest, and $\mathbf{x}$ is a vector of exogenous controls.\footnote{Because $T^*$ is binary, there is no loss of generality from writing the model in this form rather than the more familiar $y = h(T^*,\mathbf{x})+\varepsilon$. Simply define $\beta(\mathbf{x}) = h(1,\mathbf{x}) - h(0,\mathbf{x})$ and $c(\mathbf{x}) = h(0,\mathbf{x})$.}
We ask whether, and if so under what conditions, a discrete instrumental variable $z$ suffices to non-parametrically identify the causal effect $\beta(\mathbf{x})$ of $T^*$, when we observe not $T^*$ but a mis-classified binary surrogate $T$.

We proceed under the assumption of non-differential measurement error.
This condition has been widely used in the existing literature and imposes  that $T$ provides no additional information beyond that contained in $(T^*,\mathbf{x})$.
Even in this fairly standard setting, identification remains an open question: we begin by showing that the only existing identification result for this model is incorrect.
We then go on to derive the sharp identified set under the standard first-moment assumptions from the related literature.
We show that regardless of the number of values that $z$ takes on, the model is not point identified.
This motivates us to consider alternative, and slightly stronger assumptions.
We show that, given a binary instrument, the addition of a second moment independence assumption suffices to identify a model with one-sided mis-classification.
Adding a second moment restriction on the measurement error along with a third moment independence assumption for the instrument suffices to identify the model in general.
This result likewise requires only a binary $z$.

Our work relates to a large literature that considers departures from classical measurement error, by allowing the measurement error to be related to the true value of the unobserved regressor.
\cite{ChenHongTamer} obtain identification in a general class of moment condition models with mis-measured data by relying on the existence of an auxiliary dataset from which they can estimate the measurement error process.
In contrast, \cite{HuSchennach} and \cite{song2015} rely on an instrumental variable and an additional conditional location assumption on the measurement error distribution. 
More recently, \cite{HuShiuWoutersen} use a continuous instrument to identify the ratio of partial effects of two continuous regressors, one measured with error, in a linear single index model.
Unfortunately, these approaches cannot be applied to the case of a mis-measured binary regressor.

A number of papers have studied models with an exogenous binary regressor subject to non-differential measurement error.
One group of papers asks what can be learned without recourse to an instrumental variable.
An early contribution by \cite{Aigner} characterizes the asymptotic bias of OLS in this setting, and proposes a correction using outside information on the mis-classification process.
Related work by \cite{Bollinger} provides partial identification bounds.
More recently, \cite{ChenHuLewbel} use higher moment assumptions to obtain identification in a linear model, and \cite{ChenHuLewbel2} extend these results to the non-parametric setting. 
\cite{HasseltBollinger} and \cite{BollingerHasseltWP} provide additional partial identification results.
For results on the partial identification of discrete probability distributions under mis-classification, see \cite{molinari}.

Continuing under the assumption of exogeneity and non-differential measurement error, another group of papers relies on the availability of either an instrumental variable or a second measure of $T^*$.
\cite{BBS} and \cite{KRS} consider a linear model and show that when \emph{two} alternative measures $T_1$ and $T_2$ of $T^*$ are available, a non-linear GMM estimator can be used to recover the effect of interest.
Subsequently, \cite{FL} note that an instrumental variable can take the place of one of the measures.
\cite{Mahajan} extends the results of \cite{BBS} and \cite{KRS} to a more general setting using a binary instrument in place of one of the treatment measures, establishing non-parametric identification of the conditional mean function.
When $T^*$ is in fact exogenous, this coincides with the causal effect.
\cite{hu2008} derives related results when the mis-classified discrete regressor may take on more than two values.
\cite{Lewbel} provides an identification result for the same model as \cite{Mahajan} under different assumptions.
In particular, his ``instrument-like variable'' need not satisfy the usual  exclusion restriction so long as it does not interact with $T^*$ and takes on three or more values.

Much less is known about the case in which a binary, or discrete, regressor is not only mis-classified but endogenous.
The first paper to provide a formal result for this case is \cite{Mahajan}.
He extends his main result to the case of an endogenous treatment, providing an explicit proof of identification under the usual IV assumption in a model with additively separable errors.
As we show below, however, this result is false.\footnote{Appendix \ref{sec:mahajan} provides a detailed explanation of the error in \citeauthor{Mahajan}'s proof.}
Several more recent papers also consider the case of a mis-classified, endogenous, binary regressor.
\cite{kreider2012}, partially identify the effects of food stamps on health outcomes of children under weak measurement error assumptions by relying on auxiliary data.
Similarly, \cite{Batt} study the returns to schooling in a setting with multiple mis-reported measures of educational qualifications.
Unlike these two papers, our approach does not depend on the availability of auxiliary data.
In a different vein, \cite{shiu2015} uses an exclusion restriction for the participation equation and an additional valid instrument to identify the effect of a discrete, mis-classified endogenous regressor in a semi-parametric selection model.
Similarly, \cite{nguimkeu2016estimation} use exclusion restrictions for both the participation equation and measurement error equation to identify a parametric model with endogenous participation and one-sided endogenous mis-reporting. 
Unlike those of the preceding two papers, our results rely neither on parametric assumptions nor additional exclusion restrictions.
Other than \cite{Mahajan}, the paper most closely related to our own is that of \cite{Ura}, who derives partial identification results for a local average treatment effect without the non-differential assumption.
In contrast, we study an additively separable model under non-differential measurement error and derive both partial and point identification results.

The remainder of the paper is organized as follows.
Section \ref{sec:baseline} describes our model and assumptions, Section \ref{sec:ident_literature} relates our results to existing work, and Sections \ref{sec:partial}--\ref{sec:point} present our identification results.
Section \ref{sec:inference} provides a brief discussion of how to carry out inference using our identification results, and Section \ref{sec:conclusion} concludes.
Proofs appear in Appendix \ref{sec:proofs}, and we give a detailed explanation of the error in \cite{Mahajan} in Appendix \ref{sec:mahajan}.
Appendix \ref{sec:het} explains how our partial identification bounds from Section \ref{sec:partial} can be interpreted in a local average treatment effects (LATE) setting.

\section{Identification}
\label{sec:identification}

\subsection{Baseline Assumptions}
\label{sec:baseline}
As defined in the preceding section, our model is $y = c(\mathbf{x}) + \beta(\mathbf{x}) T^* + \varepsilon$, where $\varepsilon$ is a mean-zero error term, and the parameter of interest is $\beta(\mathbf{x})$ -- the effect of an unobserved, binary, endogenous regressor $T^*$.
Suppose we observe a valid and relevant binary instrument $z$.
In the discussion following Corollary \ref{cor:nonident} below, we explain how these results generalize to the case of an arbitrary discrete-valued instrument.
We assume that the model and instrument satisfy the following conditions:
\begin{assump} \mbox{}
  \label{assump:model}
  \begin{enumerate}[(i)] 
    \item $y = c(\mathbf{x}) + \beta(\mathbf{x})T^* + \varepsilon$ where $T^* \in \left\{ 0,1 \right\}$ and $\mathbb{E}[\varepsilon]=0$;
    \item  $z \in \left\{ 0,1 \right\}$, where $0 < \mathbb{P}(z=1|\mathbf{x}) < 1$, and $\mathbb{P}(T^*=1|\mathbf{x},z=1) \neq \mathbb{P}(T^*=1|\mathbf{x},z=0)$;
    \item $\mathbb{E}[\varepsilon|\mathbf{x},z] = 0$.
  \end{enumerate}
\end{assump}

Assumption \ref{assump:model}(i) is a restatement of the additively separable model from Equation \ref{eq:model}, which includes as a special case the linear model $y = c + \beta T^* + \mathbf{x}'\boldsymbol{\gamma} + \varepsilon$ that is pervasive in empirical economics. 
Assumptions \ref{assump:model}(ii) and (iii) are the textbook instrumental variable relevance and validity conditions, respectively.
Under Assumption \ref{assump:model}, the Wald estimator 
\[
  \left[\mathbb{E}\left(y|z=1,\mathbf{x}\right)-\mathbb{E}\left(y|z=0,\mathbf{x}\right)\right] / \left[ \mathbb{E}\left(T^*|z=1,\mathbf{x}\right) - \mathbb{E}\left( T^*|z=0,\mathbf{x} \right) \right]
\] 
identifies $\beta(\mathbf{x})$.
Unfortunately this estimator is infeasible, as we observe not $T^*$ but a mis-classified binary surrogate $T$.\footnote{Although it involves $T^*$, Assumption \ref{assump:model}(ii) is testable: see the discussion following Lemma \ref{lem:p_pstar}.}
To make further progress, we must impose conditions on the process that generates $T$. 
Accordingly, define the following mis-classification probabilities:
\begin{align*}
  \alpha_0(\mathbf{x},z) &= \mathbb{P}\left(T=1|T^*=0,\mathbf{x},z  \right) &
  \alpha_0(\mathbf{x}) &= \mathbb{P}\left(T=1|T^*=0,\mathbf{x}  \right)\\ 
  \alpha_1(\mathbf{x},z) &= \mathbb{P}\left(T=0|T^*=1,\mathbf{x},z  \right) &
  \alpha_1(\mathbf{x}) &= \mathbb{P}\left(T=0|T^*=1,\mathbf{x}  \right).
\end{align*}

\begin{assump} \mbox{}
  \label{assump:misclassification}
  \begin{enumerate}[(i)] 
    \item $\alpha_0(\mathbf{x},z) = \alpha_0(\mathbf{x})$,   $\alpha_1(\mathbf{x},z) = \alpha_1(\mathbf{x})$
    \item $\alpha_0(\mathbf{x}) + \alpha_1(\mathbf{x}) <1$ 
    \item $\mathbb{E}[\varepsilon|\mathbf{x},z,T^*,T] = \mathbb{E}[\varepsilon|\mathbf{x},z, T^*]$
  \end{enumerate}
\end{assump}

Assumption \ref{assump:misclassification}, or a variant thereof, is standard in the theoretical literature on mis-classification \citep{Mahajan,BBS,FL,Lewbel,hu2008} and in empirical studies that allow for measurement error in a binary or discrete variable \citep{KRS,hu2013,Batt}.
Assumption \ref{assump:misclassification} (i) states that the mis-classification probabilities do not depend on $z$.
Assumption \ref{assump:misclassification} (ii) restricts the extent of mis-classification and is equivalent to requiring that $T$ and $T^*$ be positively correlated.
Assumption \ref{assump:misclassification} (iii) is often referred to as ``non-differential measurement error.''
Intuitively, it maintains that $T$ provides no additional information about $\varepsilon$, and hence $y$, given knowledge of $(T^*,z,\mathbf{x})$.
While Assumption \ref{assump:misclassification}(ii) is quite mild, Assumptions \ref{assump:misclassification} (i) and (iii) are more restrictive, as discussed by \cite{Bound2001}.
To take a specific example, suppose that $y$ is log wage and $T^*$ is an indicator for college completion.
If $T$ is a potentially erroneous measure of college completion taken from a university's administrative records, then the assumption of non-differential measurement error is quite plausible.
If, on the other hand, $T$ is a self-report of college completion and there are ``returns to lying'' about college completion, i.e.\ employers only imperfectly observe worker ability, this assumption is less plausible.\footnote{See \cite{huLewbel} for a proposal to estimate the ``returns to lying'' in this context.}
Note, however, that our assumptions on the mis-classification process are \emph{conditional} on $\mathbf{x}$: we place no restrictions on the relationship between observed covariates and the mis-classification errors. 
In contrast, \cite{Bound2001} considers \emph{unconditional} versions of our Assumption \ref{assump:misclassification}.
Instrument validity -- Assumption \ref{assump:model} (iii) -- is more plausible after conditioning on a rich set of exogenous controls, and the same is true of our mis-classification assumptions.
For more discussion of settings in which the assumption of non-differential measurement error is warranted, see \cite{carroll2006}.

\subsection{Point Identification Results from the Literature}
\label{sec:ident_literature}

Existing results from the literature -- see for example \cite{FL} and \cite{Mahajan} -- establish that $\beta(\mathbf{x})$ is point identified if Assumptions \ref{assump:model}--\ref{assump:misclassification} are augmented to include the following condition:
\begin{assump}[Joint Exogeneity] \mbox{}
  \label{assump:jointExog}
    $\mathbb{E}[\varepsilon|\mathbf{x},z, T^*] = 0$.
\end{assump}

Assumption \ref{assump:jointExog} strengthens the mean independence condition from Assumption \ref{assump:model} (iii) to hold \emph{jointly} for $T^*$ and $z$.
By iterated expectations, this implies that $T^*$ is exogenous, i.e.\ $\mathbb{E}[\varepsilon|\mathbf{x},T^*] = 0$.
If $T^*$ is endogenous, Assumption \ref{assump:jointExog} clearly fails.
\cite{Mahajan} argues, however, that the following restriction, along with our Assumptions \ref{assump:model}--\ref{assump:misclassification}, suffices to identify $\beta(\mathbf{x})$ when $T^*$ may be endogenous:
\begin{assump}[\cite{Mahajan} Equation 11] \mbox{}
  \label{assump:mahajan}
  $\mathbb{E}[\varepsilon|\mathbf{x}, z, T^*, T] = \mathbb{E}[\varepsilon|\mathbf{x},T^*]$.
\end{assump}
Assumption \ref{assump:mahajan} does not require $\mathbb{E}[\varepsilon|\mathbf{x},T^*]$ to be zero, but maintains that it does not vary with $z$.
We show in Appendix \ref{sec:mahajan}, however, that under 
Assumptions \ref{assump:model}--\ref{assump:misclassification}, Assumption \ref{assump:mahajan} can only hold if $T^*$ is exogenous.
If $z$ is a valid instrument and $T^*$ is endogenous, then Assumption \ref{assump:mahajan} implies that there is no first-stage relationship between $z$ and $T^*$.
As such, identification in the case where $T^*$ is endogenous is an open question.


\subsection{Partial Identification}
\label{sec:partial}

In this section we derive the sharp identified set under Assumptions \ref{assump:model}--\ref{assump:misclassification} and show that  $\beta(\mathbf{x})$ is not point identified.
For a discussion of how our partial identification results can be interpreted in a local average treatment effects (LATE) setting, see Appendix \ref{sec:het}.

To simplify the notation, define the following shorthand for the unobserved and observed first stage probabilities
\begin{equation}
  p^*_k(\mathbf{x}) = \mathbb{P}(T^*=1|\mathbf{x},z=k), \quad
  p_k(\mathbf{x}) = \mathbb{P}(T=1|\mathbf{x},z=k).
  \label{eq:pk_def}
\end{equation}
We first state two lemmas that that will be used repeatedly below.
\begin{lem}
\label{lem:p_pstar}
  Under Assumption \ref{assump:misclassification} (i),
\begin{align*}
  \left[ 1 - \alpha_0(\mathbf{x}) - \alpha_1(\mathbf{x}) \right]p^*_k(\mathbf{x}) &= p_k(\mathbf{x}) - \alpha_0(\mathbf{x})\\
  \left[ 1 - \alpha_0(\mathbf{x}) - \alpha_1(\mathbf{x}) \right]\left[1 - p^*_k(\mathbf{x}) \right]&= 1 - p_k(\mathbf{x}) - \alpha_1(\mathbf{x})
\end{align*}
where the first-stage probabilities $p_k^*(\mathbf{x})$ and $p_k(\mathbf{x})$ are as defined in Equation \ref{eq:pk_def}.
\end{lem} 

\begin{lem}
  \label{lem:wald}
  Under Assumptions \ref{assump:model} and \ref{assump:misclassification} (i)--(ii), $$\beta(\mathbf{x}) \mbox{Cov}(z,T|\mathbf{x}) = \left[ 1 - \alpha_0(\mathbf{x}) - \alpha_1(\mathbf{x}) \right]\mbox{Cov}(y,z|\mathbf{x})$$
\end{lem}

Lemma \ref{lem:p_pstar} relates the observed first-stage probabilities $p_k(\mathbf{x})$ to their unobserved counterparts $p^*_k(\mathbf{x})$ in terms of the mis-classification probabilities $\alpha_0(\mathbf{x})$ and $\alpha_1(\mathbf{x})$.
By Assumption \ref{assump:misclassification} (ii), $1 - \alpha_0(\mathbf{x}) - \alpha_1(\mathbf{x}) > 0$ so that Lemma \ref{lem:p_pstar} bounds $\alpha_0(\mathbf{x})$ and $\alpha_1(\mathbf{x})$ in terms of the observed first-stage probabilities.
Moreover, by taking differences evaluated at $k=1$ and $k=0$, this Lemma shows that $p_0^*(\mathbf{x}) = p_1^*(\mathbf{x})$ if and only if $p_0(\mathbf{x})=p_1(\mathbf{x})$.
In other words, Assumption \ref{assump:model} (ii) is testable under Assumption \ref{assump:misclassification} (ii).
Lemma \ref{lem:wald} relates the instrumental variables (IV) estimand, $\mbox{Cov}(y,z|\mathbf{x})/\mbox{Cov}(z,T|\mathbf{x})$, to the mis-classification probabilities.
Since $1 - \alpha_0(\mathbf{x}) - \alpha_1(\mathbf{x}) > 0$, IV is biased \emph{upwards} in the presence of mis-classification.
Together these lemmas bound the causal effect of interest: $\beta(\mathbf{x})$ lies between the reduced form and IV estimators.
Without Assumption \ref{assump:misclassification} (iii), non-differential measurement error, these bounds are sharp.

\begin{thm}
  Under Assumptions \ref{assump:model} and \ref{assump:misclassification} (i)--(ii), $\alpha_0(\mathbf{x}) \leq p_k(\mathbf{x}) \leq 1 -  \alpha_1(\mathbf{x})$ for  $k = 0, 1$ and
  \begin{equation}
    \mathbb{E}[y|\mathbf{x},z=k] = c(\mathbf{x}) + \beta(\mathbf{x}) \left[\frac{p_k(\mathbf{x}) - \alpha_0(\mathbf{x})}{1 - \alpha_0(\mathbf{x}) - \alpha_1(\mathbf{x})}\right].
    \label{eq:identsetI}
  \end{equation}
Provided that $p_0(\mathbf{x}) \neq p_1(\mathbf{x})$, these expressions characterize the sharp identified set for $c(\mathbf{x})$, $\beta(\mathbf{x})$, $\alpha_0(\mathbf{x})$, and $\alpha_1(\mathbf{x})$.
  \label{thm:sharpI}
\end{thm}

\begin{cor}
  Under the conditions of Theorem \ref{thm:sharpI}, the sharp identified set for $\beta(\mathbf{x})$ is the closed interval between the reduced form estimand $\mbox{Cov}(y,z|\mathbf{x})/\mbox{Var}(z|\mathbf{x})$ and the IV estimand $\mbox{Cov}(y,z|\mathbf{x})/\mbox{Cov}(z,T|\mathbf{x})$.
\label{cor:sharpBeta1}
\end{cor}

Corollary \ref{cor:sharpBeta1} follows by taking differences of the expression for $\mathbb{E}[y|\mathbf{x},z=k]$ across $k=1$ and $k=0$, and substituting the maximum and minimum value for $\alpha_0(\mathbf{x}) + \alpha_1(\mathbf{x})$ consistent with the observed first-stage probabilities.\footnote{If \emph{a priori} restrictions on $\alpha_0$ and $\alpha_1$ are available, e.g.\ $\alpha_0 = 0$, $\alpha_1 = 0$, or $\alpha_0 = \alpha_1$, these bounds can be improved. For more discussion, see Corollary 2.2 of \cite{DiTragliaGarciaWP2017}.}
Note that the only role of the condition $p_0(\mathbf{x}) \neq p_1(\mathbf{x})$ in the preceding two results is to ensure that it is possible to satisfy Assumption \ref{assump:model} (ii).
\cite{FL} point out that the IV estimand provides an upper bound for $\beta(\mathbf{x})$, and Lemmas \ref{lem:p_pstar}--\ref{lem:wald} are well-known in the literature \citep[see e.g.][]{FL,Mahajan}.
Nevertheless, we are unaware of any published result that explicitly states both bounds from Corollary \ref{cor:sharpBeta1} or proves that they are sharp under Assumptions \ref{assump:model} and \ref{assump:misclassification} (i)--(ii).

Neither Theorem \ref{thm:sharpI} nor Corollary \ref{cor:sharpBeta1} imposes Assumption \ref{assump:misclassification} (iii) -- non-differential measurement error.
While this assumption plays an important role in existing identification results for an exogenous $T^*$ (see Section \ref{sec:ident_literature}), its identifying power under endogeneity has not been addressed in the literature.\footnote{The only exception is the incorrect result of \cite{Mahajan} described in Section \ref{sec:ident_literature} and Appendix \ref{sec:mahajan}.}
We now show that this assumption in general yields further restrictions on probabilities $\alpha_0(\mathbf{x})$ and $\alpha_1(\mathbf{x})$, but fails to point identify $\beta(\mathbf{x})$.
To simplify the proof of sharpness, we assume that $y$ is continuously distributed, which is natural in an additively separable model.
Without this assumption, the bounds that we derive are still valid, but may not be sharp. 
Nevertheless, the reasoning from our proof can be generalized to cases in which $y$ does not have a continuous support set.

\begin{thm}
  \label{thm:sharpII}
  Suppose that the conditional distribution of $y$ given $(\mathbf{x}, T,z)$ is continuous.
  Further suppose that the conditions of Theorem \ref{thm:sharpI} and Assumption \ref{assump:misclassification} (iii) hold.
  For any $k$ such that $\mathbb{E}\left[ y|\mathbf{x},T=0,z=k \right] \neq \mathbb{E}\left[ y|\mathbf{x},T=1,z=k \right]$, let $\mathcal{A}_k$ denote the set of pairs $\big(\alpha_0(\mathbf{x}), \alpha_1(\mathbf{x}) \big)$ such that  $\alpha_0(\mathbf{x}) < p_k(\mathbf{x}) < 1 -  \alpha_1(\mathbf{x})$ and
\[
  \underline{\mu}_{tk}\bigg( \underline{q}_{tk}\big( \alpha_0(\mathbf{x}), \alpha_1(\mathbf{x}), \mathbf{x}\big) , \,\mathbf{x} \bigg)\leq 
  \mu_{k}\big( \alpha_0(\mathbf{x}),\mathbf{x} \big)\leq 
  \overline{\mu}_{tk}\bigg(\overline{q}_{tk}\big( \alpha_0(\mathbf{x}), \alpha_1(\mathbf{x}), \mathbf{x}\big), \,\mathbf{x} \bigg)
\]
for all $t = 0,1$ where
\begin{align*}
  \underline{\mu}_{tk}\big( q,\mathbf{x} \big) = \mathbb{E}\left[ y\left|\right.y\leq q, \mathbf{x},T=t, z=k\right], \quad \quad
  \overline{\mu}_{tk}\big(q,\mathbf{x} \big) = \mathbb{E}\left[ y\left|\right. y > q, \mathbf{x}, T=t, z=k\right]
\end{align*}
  \begin{align*}
  \mu_k\big(\alpha_0(\mathbf{x}),\mathbf{x}\big) &= 
  \frac{p_k(\mathbf{x}) \mathbb{E}[y|\mathbf{x},z=k,T=1] - \alpha_0(\mathbf{x}) \mathbb{E}[y|\mathbf{x},z=k]}{p_k(\mathbf{x}) - \alpha_0(\mathbf{x})}
\end{align*}
and we define 
\begin{align*}
  \underline{q}_{tk}\big(\alpha_0(\mathbf{x}),\alpha_1(\mathbf{x}),\mathbf{x}\big) &= F^{-1}_{tk}\bigg(r_{tk}\big(\alpha_0(\mathbf{x}),\alpha_1(\mathbf{x}), \mathbf{x}\big)\, \bigg|\,\mathbf{x}\bigg)\\
  \overline{q}_{tk}\big(\alpha_0(\mathbf{x}),\alpha_1(\mathbf{x}),\mathbf{x}\big) &= F^{-1}_{tk}\bigg(1 - r_{tk}\big(\alpha_0(\mathbf{x}), \alpha_1(\mathbf{x}),\mathbf{x}\big) \,\bigg|\,\mathbf{x}\bigg)
\end{align*}
where $F_{tk}^{-1}(\cdot|\mathbf{x})$ is the conditional quantile function of $y$ given $(\mathbf{x},T=t,z=k)$,  
\begin{align*}
  r_{0k}\big(\alpha_0(\mathbf{x}),\alpha_1(\mathbf{x}),\mathbf{x}\big) &= \frac{\alpha_1(\mathbf{x})}{1 - p_k(\mathbf{x})} \left[ \frac{p_k(\mathbf{x}) - \alpha_0(\mathbf{x})}{1 - \alpha_0(\mathbf{x}) - \alpha_1(\mathbf{x})} \right]\\
  r_{1k}\big(\alpha_0(\mathbf{x}),\alpha_1(\mathbf{x}),\mathbf{x}\big) &= \frac{1 - \alpha_1(\mathbf{x})}{p_k(\mathbf{x})} \left[ \frac{p_k(\mathbf{x}) - \alpha_0(\mathbf{x})}{1 - \alpha_0(\mathbf{x}) - \alpha_1(\mathbf{x})} \right]
\end{align*}
and $p_k(\mathbf{x})$ is defined in Equation \ref{eq:pk_def}.
The sharp identified set for $c(\mathbf{x})$, $\beta(\mathbf{x})$, $\alpha_0(\mathbf{x})$ and $\alpha_1(\mathbf{x})$ is characterized by Equation
\ref{eq:identsetI} and $\big(\alpha_0(\mathbf{x}), \alpha_1(\mathbf{x})\big) \in \mathcal{A}^*$ where
\begin{enumerate}[(i)]
  \item $\mathcal{A}^* \equiv \mathcal{A}_0 \cap \mathcal{A}_1$ if $\mathbb{E}[y|\mathbf{x},T=0,z=k] \neq \mathbb{E}[y|\mathbf{x},T=1,z=k]$ for all $k=0,1$;
  \item $\mathcal{A}^* \equiv \mathcal{A}_k$ if $\mathbb{E}[y|\mathbf{x},T=0,z=k] \neq \mathbb{E}[y|\mathbf{x},T=1,z=k]$ and $\mathbb{E}[y|\mathbf{x},T=0,z=\ell] = \mathbb{E}[y|\mathbf{x},T=1,z=\ell]$;
  \item $\mathcal{A}^* \equiv \left\{ \big(\alpha_0(\mathbf{x}), \alpha_1(\mathbf{x})\big)\colon \alpha_0(\mathbf{x}) \leq p_k(\mathbf{x}) \leq 1 - \alpha_1(\mathbf{x}) \mbox{ for all } k \right\}$ if $\mathbb{E}[y|\mathbf{x},T=0,z=k] = \mathbb{E}[y|\mathbf{x},T=1,z=k]$ for all $k=0,1$.
\end{enumerate}
\end{thm}


Imposing Assumption \ref{assump:misclassification} (iii) strictly improves upon the identified set from Theorem \ref{thm:sharpI} unless $\mathbb{E}[y|\mathbf{x},T=0,z=k]=\mathbb{E}[y|\mathbf{x},T=1,z=k]$ for all $k$. 
Even if $\beta(\mathbf{x}) = 0$, the difference of these observable means is generically nonzero.\footnote{Suppress dependence on $\mathbf{x}$ for simplicity.
There are only two settings in which $\mathbb{E}[y|T=0,z=k] = \mathbb{E}[y|T=1,z=k]$.
The first is if the true value of either $\alpha_0$ or $\alpha_1$ lies at the upper boundary of the identified set from Theorem \ref{thm:sharpI}.
The second is if $\beta = \mathbb{E}[\varepsilon|T^*=0,z=k] - \mathbb{E}[\varepsilon|T^*=0,z=k]$.}
The intuition for Theorem \ref{thm:sharpII} is as follows.
For simplicity, suppress dependence on $\mathbf{x}$.
Now, fix $(T=t, z=k)$ and $(\alpha_0, \alpha_1)$.
The observed distribution of $y$ given $(T=t,z=k)$, call it $F_{tk}$, is a mixture of two unobserved distributions: the distribution of $y$ given $(T=k,z=k,T^*=1)$, call it $F^1_{tk}$, and the distribution of $y$ given $(T=t,z=k,T^*=0)$, call it $F^{0}_{tk}$.
The mixing probabilities are $r_{tk}$ and $1-r_{tk}$ from the statement of Theorem \ref{thm:sharpII} and are fully determined by $(\alpha_0, \alpha_1)$ and $p_k$.
Assumptions \ref{assump:model} (i) and \ref{assump:misclassification} (ii) imply that the unobserved means $\mathbb{E}[y|T^*,T,z]$  are fully determined by $(\alpha_0, \alpha_1)$ given the observed means $\mathbb{E}[y|T,z]$.
The question is whether it is possible, given the observed distribution $F_{tk}$, to construct $F^1_{tk}$ and $F^{0}_{tk}$ with the required values for $\mathbb{E}[y|T^*,T,z]$ such that $F_{tk} = r_{tk} F^{1}_{tk} + (1 - r_{tk}) F^{0}_{tk}$ for all combinations $(t,k)$. 
If not, then $(\alpha_0, \alpha_1)$ does not belong to the identified set.
Our proof provides necessary and sufficient conditions for such a mixture to exist at a given point $(\alpha_0, \alpha_1)$.
We can then appeal to the reasoning from Theorem \ref{thm:sharpI} to complete the argument.
By ruling out values for $\alpha_0$ and $\alpha_1$, Theorem \ref{thm:sharpII} restricts $\beta$ via Lemma \ref{lem:wald}. 
While these restrictions can be very informative,  they do not yield point identification.

\begin{cor}
  Under Assumptions \ref{assump:model} and \ref{assump:misclassification} the identified set for $\beta(\mathbf{x})$ contains both the IV estimand $\mbox{Cov}(y,z|\mathbf{x})/\mbox{Cov}(z,T|\mathbf{x})$ and the true coefficient $\beta(\mathbf{x})$.
  \label{cor:nonident}
\end{cor}

Corollary \ref{cor:nonident} follows by Lemma \ref{lem:wald} because $\alpha_0(\mathbf{x})=\alpha_1(\mathbf{x})=0$ always belongs to the sharp identified set from Theorem \ref{thm:sharpII}.
Non-differential measurement error cannot exclude the possibility that there is no mis-classification because in this case it is trivial to construct the required mixtures.
Although we focus throughout this paper on the case of a binary instrument, one might wonder whether point identification can be achieved by increasing the support of $z$, perhaps along the lines of \cite{Lewbel}.
The answer turns out to be no.
Suppose that we were to modify Assumptions \ref{assump:model} and \ref{assump:misclassification} to hold for all values of $z$ in some discrete support set.
By Lemma \ref{lem:wald}, a binary instrument identifies $\beta(\mathbf{x})$ up to knowledge of the mis-classification probabilities $\alpha_0(\mathbf{x})$ and $\alpha_1(\mathbf{x})$.
It follows that \emph{any} pair of values $(k,\ell)$ in the support set of $z$ identifies the same object.
Accordingly, to identify $\beta(\mathbf{x})$ it is necessary and sufficient to identify the mis-classification probabilities.
A binary instrument fails to identify these probabilities because we can never exclude the possibility of zero mis-classification.
The same is true of a discrete $K$-valued instrument. 
Increasing the support of $z$ does, however, shrink the identified set by increasing the number of restrictions available: in this case Theorems \ref{thm:sharpI}--\ref{thm:sharpII} continue to apply replacing ``$k=0,1$'' with ``for all $k$.''


\subsection{Point Identification}
\label{sec:point}
The results of the preceding section establish that $\beta(\mathbf{x})$ is not point identified under Assumptions \ref{assump:model} and \ref{assump:misclassification}.
In light of this, there are two possible ways to proceed: either one can report partial identification bounds based on our characterization of the sharp identified set from Theorem \ref{thm:sharpII}, or one can attempt to impose stronger assumptions to obtain point identification.
In this section we consider the second possibility.
We begin by defining the following functions of the model parameters: 
\begin{align}
  \label{eq:theta1_def}
  \theta_1(\mathbf{x}) &= \beta(\mathbf{x})\left[ 1 -  \alpha_0(\mathbf{x}) - \mathbf{\alpha}_1(\mathbf{x}) \right]^{-1}\\
  \label{eq:theta2_def}
  \theta_2(\mathbf{x}) &= \left[\theta_1(\mathbf{x})\right]^2 \left[ 1 + \alpha_0(\mathbf{x}) - \alpha_1(\mathbf{x})\right] \\
  \label{eq:theta3_def}
  \theta_3(\mathbf{x}) &= \left[\theta_1(\mathbf{x})\right]^3\left[ \left\{ 1 -\alpha_0(\mathbf{x}) - \alpha_1(\mathbf{x}) \right\}^2 + 6\alpha_0(\mathbf{x})\left\{ 1 - \alpha_1(\mathbf{x}) \right\} \right]
\end{align}
Now consider the following additional assumption:
\begin{assump} \mbox{}
  \label{assump:2ndMoment}
    $\mathbb{E}[\varepsilon^2|\mathbf{x},z] = \mathbb{E}[\varepsilon^2|\mathbf{x}]$ 
\end{assump}
Assumption \ref{assump:2ndMoment} is a \emph{second moment} version of the standard mean exclusion restriction for the instrument $z$ -- Assumption \ref{assump:model} (iii).
It requires that the conditional variance of the error term given the covariates $\mathbf{x}$ does not depend on $z$, but does \emph{not} require homoskedasticity with respect to $\mathbf{x}, T^*$ or $T$.
Assumption \ref{assump:2ndMoment} allows us to derive the following lemma:

\begin{lem} Under Assumptions \ref{assump:model}, \ref{assump:misclassification} and \ref{assump:2ndMoment}, 
\[
  \mbox{Cov}(y^2,z|\mathbf{x}) = 2\mbox{Cov}(yT,z|\mathbf{x}) \theta_1(\mathbf{x}) -\mbox{Cov}(T,z|\mathbf{x})\theta_2(\mathbf{x}) 
\]
  where $\theta_1(\mathbf{x})$ and $\theta_2(\mathbf{x})$ are defined in Equations \ref{eq:theta1_def}--\ref{eq:theta2_def}.
  \label{lem:eta2}
\end{lem}

Lemma \ref{lem:wald} identifies $\theta_1(\mathbf{x})$.
Since $\mbox{Cov}(z,T|\mathbf{x}) \neq 0$ by Assumption \ref{assump:model} (ii), we can solve for $\theta_2(\mathbf{x})$ in terms of observables only, using Lemma \ref{lem:eta2}.
Given knowledge of $\theta_1(\mathbf{x})$, we can solve Equation \ref{eq:theta2_def} for the difference of mis-classification rates so long as $\beta(\mathbf{x}) \neq 0$.

\begin{cor}
  Under Assumptions \ref{assump:model}--\ref{assump:misclassification} and \ref{assump:2ndMoment}, $\alpha_1(\mathbf{x}) - \alpha_0(\mathbf{x})$ is identified so long as $\beta(\mathbf{x}) \neq 0$.
  \label{cor:alpha_diff}
\end{cor}
Corollary \ref{cor:alpha_diff} identifies the difference of mis-classification error rates.
Hence, under one-sided mis-classification, $\alpha_0(\mathbf{x}) = 0$ or $\alpha_1(\mathbf{x}) = 0$, augmenting our baseline Assumptions \ref{assump:model}--\ref{assump:misclassification} with Assumption \ref{assump:2ndMoment} suffices to identify $\beta(\mathbf{x})$.
Notice that $\beta(\mathbf{x})=0$ if and only if $\theta_1(\mathbf{x}) = 0$.
Thus, $\beta(\mathbf{x})$ is still identified in the case where Corollary \ref{cor:alpha_diff} fails to apply.

Assumption \ref{assump:2ndMoment} does not suffice to identify $\beta(\mathbf{x})$ without \emph{a priori} restrictions on the mis-classification error rates.
To achieve identification in the general case, we impose the following additional conditions:
\begin{assump} \mbox{}
  \label{assump:3rdMoment}
  \begin{enumerate}[(i)] 
    \item $\mathbb{E}[\varepsilon^2|\mathbf{x},z,T^*,T] = \mathbb{E}[\varepsilon^2|\mathbf{x},z, T^*]$
    \item $\mathbb{E}[\varepsilon^3|\mathbf{x},z] = \mathbb{E}[\varepsilon^3|\mathbf{x}]$
  \end{enumerate}
\end{assump}

Assumption \ref{assump:3rdMoment} (i) is a second moment version of the non-differential measurement error assumption, Assumption \ref{assump:misclassification} (iii).
It requires that, given knowledge of $(\mathbf{x}, T^*,z)$, $T$ provides no additional information about the variance of the error term.
Note that Assumption \ref{assump:3rdMoment} (i) does not require homoskedasticity of $\varepsilon$ with respect to $\mathbf{x}$ or $T^*$.
Assumption \ref{assump:3rdMoment} (ii) is a third moment version of Assumption \ref{assump:2ndMoment}.
It requires that the conditional third moment of the error term given $\mathbf{x}$ does not depend on $z$.
This condition neither requires nor excludes skewness in the error term conditional on covariates: it merely states that the skewness is unaffected by the instrument.
While Assumptions \ref{assump:2ndMoment} and \ref{assump:3rdMoment} may appear somewhat unusual, they are implied by the more intuitive independence conditions
$\varepsilon \indep z |\mathbf{x}$ and $\varepsilon \indep T | (\mathbf{x}, T^*, z)$.
Although $\mathbb{E}[\varepsilon|\mathbf{x},z]=0$ and $\mathbb{E}[\varepsilon|\mathbf{x},z,T^*,T] = \mathbb{E}[\varepsilon|\mathbf{x},z,T^*]$ are technically weaker than assuming full independence, we would be somewhat dubious of any supposed ``natural experiment'' that purportedly satisfied mean exclusion but not independence.
Indeed, as discussed by \cite{ImbensRubin1997}, an instrument satisfying mean exclusion but not independence could become invalid if the outcome variable were transformed, for example by taking logs.
As it is not uncommon for applied papers to report results in both logs and levels \citep[e.g.][]{angrist1990}, our view is that researchers \emph{implicitly} assume more than mean exclusion in typical applications of instrumental variables. 
Analogous reasoning applies to the non-differential measurement error assumption.

Assumption \ref{assump:3rdMoment} allows us to derive the following Lemma which, combined with Lemma \ref{lem:eta2}, leads to point identification: 

\begin{lem}
  Under Assumptions \ref{assump:model}--\ref{assump:misclassification} and \ref{assump:2ndMoment}--\ref{assump:3rdMoment}, 
\[
    \mbox{Cov}(y^3,z|\mathbf{x}) = 3 \mbox{Cov}(y^2T,z|\mathbf{x}) \theta_1(\mathbf{x}) -3\mbox{Cov}(yT,z|\mathbf{x}) \theta_2(\mathbf{x}) + \mbox{Cov}(T,z|\mathbf{x}) \theta_3(\mathbf{x})
\]
where $\theta_1(\mathbf{x}),\theta_2(\mathbf{x})$ and $\theta_3(\mathbf{x})$ are defined in Equations \ref{eq:theta1_def}--\ref{eq:theta2_def}.
\label{lem:eta3}
\end{lem}

\begin{thm}
  Under Assumptions \ref{assump:model}--\ref{assump:misclassification} and \ref{assump:2ndMoment}--\ref{assump:3rdMoment}, $\beta(\mathbf{x})$ is identified.
  If $\mathbf{\beta}(\mathbf{x}) \neq 0$, then $\alpha_0(\mathbf{x})$ and $\alpha_1(\mathbf{x})$ are likewise identified.
  \label{thm:main_ident}
\end{thm}

Lemmas \ref{lem:wald}--\ref{lem:eta3} yield a linear system of three equations in $\theta_1(\mathbf{x}), \theta_2(\mathbf{x})$ and $\theta_3(\mathbf{x})$.
Under Assumption \ref{assump:model} (ii), the system has a unique solution so $\theta_1(\mathbf{x}), \theta_2(\mathbf{x})$ and $\theta_3(\mathbf{x})$ are identified.
The proof of Theorem \ref{thm:main_ident} shows that, so long as $\beta(\mathbf{x})\neq 0$, Equations \ref{eq:theta1_def}--\ref{eq:theta3_def} can be solved for $\beta(\mathbf{x})$, $\alpha_0(\mathbf{x})$ and $\alpha_1(\mathbf{x})$.
In particular, using steps from the proof of Theorem \ref{thm:main_ident}
\[
  \beta(\mathbf{x}) = \mbox{sign}\big[ \theta_1(\mathbf{x}) \big] \sqrt{3 \big[ \theta_2(\mathbf{x})/\theta_1(\mathbf{x}) \big]^2 - 2\big[\theta_3(\mathbf{x})/\theta_1(\mathbf{x})\big]}.
\]
If we relax Assumption \ref{assump:misclassification} (ii) and assume $\alpha_0(\mathbf{x}) + \alpha_1(\mathbf{x}) \neq 1$ only, $\beta(\mathbf{x})$ is only identified up to sign: in this case the sign of $\theta_1(\mathbf{x})$ need not equal that of $\beta(\mathbf{x})$.

\section{Estimation and Inference}
\label{sec:inference}
We now briefly outline how the identification results from Section \ref{sec:identification} can be used to estimate and carry out statistical inference for the parameters of interest: $\big(\alpha_0(\mathbf{x}), \alpha_1(\mathbf{x}), \beta(\mathbf{x})\big)$.
Lemmas \ref{lem:wald}--\ref{lem:eta3} yield a system of linear moment equations in the reduced form parameters $\boldsymbol{\theta}'(\mathbf{x}) = \big(\theta_1(\mathbf{x}), \theta_2(\mathbf{x}),\theta_3(\mathbf{x})\big)$.
Defining a vector of intercepts $\boldsymbol{\kappa}'(\mathbf{x}) = \big(\kappa_1(\mathbf{x}), \kappa_2(\mathbf{x}), \kappa_3(\mathbf{x})\big)$,
and a vector of observables $\mathbf{w}' = (T, y, yT, y^2, y^2 T, y^3)$, we can write this system as
\begin{align}
&\mathbb{E}\left[
  \bigg\{\boldsymbol{\Psi}\big(\boldsymbol{\theta}(\mathbf{x})\big)\mathbf{w}_i - \boldsymbol{\kappa}(\mathbf{x})\bigg\} \otimes 
\left(
\begin{array}{c}
  1 \\ z
\end{array}\right)\Bigg| \mathbf{x} = \boldsymbol{x}
\right] = \mathbf{0}
\label{eq:MCs_endog}\\
  &\boldsymbol{\Psi}\big(\boldsymbol{\theta}(\mathbf{x})\big) \equiv 
  \left[
  \begin{array}{rrrrrr}
    -\theta_1(\mathbf{x}) & 1 & 0 & 0 & 0 & 0\\
    \theta_2(\mathbf{x}) & 0 & -2\theta_1(\mathbf{x}) & 1 & 0 & 0\\ 
    -\theta_3(\mathbf{x}) & 0 & 3\theta_2(\mathbf{x}) & 0 & -3\theta_1(\mathbf{x}) & 1
\end{array}\right].
\end{align}
Using Equations \ref{eq:theta1_def}--\ref{eq:theta3_def}, we can re-write  $\mathbf{\Psi}$ as a function of $\big(\alpha_0(\mathbf{x}), \alpha_1(\mathbf{x}), \beta(\mathbf{x})\big)$, leaving us with a just-identified, non-parametric conditional moment problem.
Because the conditioning variables in Equation \ref{eq:MCs_endog} are the same as the arguments of the unknown functions $(\alpha_0, \alpha_1, \beta)$, this problem fits within the framework of \cite{Lewbel2007}, permitting straightforward estimation and inference via a local GMM procedure. 
If $\beta(\mathbf{x})$ is close to zero, however, this procedure can perform poorly; in this case the moment conditions from Equations \ref{eq:MCs_endog}, are only weakly informative about $\alpha_0(\mathbf{x})$ and $\alpha_1(\mathbf{x})$.
An earlier version of this paper \citep{DiTragliaGarciaWP2017} discusses this problem in more detail and provides a solution based on generalized moment selection \citep{AndrewsSoares} that combines the moment inequalities implied by our partial identification results from Section \ref{sec:partial} with the moment equalities from Equation \ref{eq:MCs_endog}.

\section{Conclusion}
\label{sec:conclusion}

This paper has studied identification and inference for a mis-classified, binary, endogenous regressor in an additively separable model using a discrete instrumental variable.
We have shown that the only existing identification result for this model is incorrect, and gone on to derive the sharp identified set under standard first-moment assumptions from the literature.
Strengthening these assumptions to hold for second and third moments, we have established point identification for the effect of interest.
An interesting extension of the results presented above would be to consider the case of discrete regressors that take on more than two values.

\appendix
\numberwithin{equation}{section}
\numberwithin{table}{section}
\numberwithin{figure}{section}
\singlespacing \small
\section{Proofs}
\label{sec:proofs}
All of the results in this paper hold $\mathbf{x}$ \emph{fixed}.
This allows us to completely ignore the presence of covariates in the proofs that follow.
Accordingly we work in terms of scalars $\alpha_0, \alpha_1, \beta, p_k$, etc.\ rather than functions $\alpha(\mathbf{x})$, $\alpha_1(\mathbf{x})$, $\beta(\mathbf{x}), p_k(\mathbf{x})$.
The former should be understood as the value of the latter evaluated at some particular $\mathbf{x}$.

\subsection{Partial Identification Results}
\begin{proof}[Proof of Lemma \ref{lem:p_pstar}]
  Follows from a simple calculation using the law of total probability.
\end{proof}

\begin{proof}[Proof of Lemma \ref{lem:wald}]
  Immediate since $\mbox{Cov}(z,T) = (1 - \alpha_0 - \alpha_1) \mbox{Cov}(z,T^*)$ by Lemma \ref{lem:p_pstar}.
\end{proof}

\begin{proof}[Proof of Theorem \ref{thm:sharpI}]
  To show that $\alpha_0 \leq p_k \leq 1 - \alpha_1$,  substitute $p_k^*=0$ and $p_k^*=1$, respectively, into Lemma \ref{lem:p_pstar} and rearrange.
  To show that $\mathbb{E}[y|z=k] = c + \beta(p_k - \alpha_0) / (1 - \alpha_0 - \alpha_1)$, take conditional expectations of Equation \ref{eq:model} and apply Assumption \ref{assump:model} (iii) and Lemma \ref{lem:p_pstar}.

To prove sharpness we need to show that for any $(c,\beta, \alpha_0, \alpha_1)$ that satisfy $\alpha_0 \leq p_k \leq 1 - \alpha_1$ and $\mathbb{E}[y|z=k] = c + \beta(p_k - \alpha_0)/(1 - \alpha_0 - \alpha_1)$  we can construct a valid joint distribution for $(y,T,T^*,z)$ that is compatible with the observed distribution of $(y,T,z)$, provided that $p_1 \neq p_0$.
To establish this result, we factorize the joint distribution of $(y,T,T^*,z)$ into the product of a conditional $y|(T,T^*,z)$ and marginal $(T,T^*,z)$.
The argument proceeds in two steps.
Our first step relies on the fact that Assumptions \ref{assump:model} (i) and (iii) do not constrain the distribution of $(T,T^*,z)$ while \ref{assump:model} (ii) and \ref{assump:misclassification} (i)--(ii) constrain \emph{only} the distribution of $(T,T^*,z)$.
Under these latter three assumptions, we show how to construct a valid joint distribution for $(T,T^*,z)$ that is compatible with the observed distribution of $(T,z)$ for any $(\alpha_0,\alpha_1)$ satisfying $\alpha_0 \leq p_k \leq 1 - \alpha_1$.
Our second step shows how to construct a valid conditional distribution for $y$ given $(T,T^*,z)$ under Assumptions \ref{assump:model} (i) and (iii) that is compatible with the observed conditional distribution of $y$ given $(T,z)$ for any $(c,\beta,\alpha_0, \alpha_1)$ satisfying $\mathbb{E}[y|z=k] = c + \beta (p_k - \alpha_0)(1 - \alpha_0 - \alpha_1)$.
Combining the two steps gives the required joint distribution for $(y,T^*,T,z)$.

For the first step, we need to construct a valid joint probability mass function $p(T^*,T,z)$ with support set $\left\{ 0,1 \right\} \times \left\{ 0,1 \right\} \times \left\{ 0,1 \right\}$.
 By Assumption \ref{assump:misclassification} (i), $p(T|T^*,z) = p(T|T^*)$ and hence 
\[
  p(T^*,T,z) = p(T|T^*)p(T^*|z)p(z).
\]
Since $p(z)$ is observed, to construct a valid joint probability mass function $p(T^*,T,z)$ it suffices to construct valid \emph{conditional} probability mass functions $p(T|T^*)$ and $p(T^*|z)$.
Since $\alpha_0 \leq p_k \leq 1 - \alpha_1$, both $\alpha_0$ and $\alpha_1$ are guaranteed to lie between zero and one.
This gives a valid construction of $p(T|T^*)$.
Moreover the corresponding values of $p_k^*$ implied by Lemma \ref{lem:p_pstar} are also guaranteed to lie between zero and one.
This gives a valid construction of $p(T^*|z)$ that satisfies Assumption \ref{assump:model} (ii), since $p_1 \neq p_0$ by assumption and $(p_1 - p_0) = (p_1^* - p_0^*)(1 - \alpha_0 - \alpha_1)$ by Lemma \ref{lem:p_pstar}.
Because our construction relies on Lemma \ref{lem:p_pstar}, which is simply an application of the law of total probability, the resulting distribution $p(T,T^*,z)$ is automatically compatible with $p(T,z) = p(T|z)p(z)$.

For the second step, we need to construct a valid conditional distribution for $y$ given $(T,T^*,z)$.
To begin we define the following notation:
\begin{align*}
r_{tk} &\equiv \mathbb{P}(T^*=1|T=t,z=k) &
F_{t}(\tau) &\equiv \mathbb{P}(y \leq \tau|z=k) \\
F_{tk}(\tau) &\equiv \mathbb{P}(y \leq \tau|T=t, z=k) & 
F_{tk}^{t^*}(\tau) &\equiv \mathbb{P}(y \leq \tau|T^*=t^*,T=t, z=k)\\
G_k(\tau) &\equiv \mathbb{P}(\varepsilon \leq \tau|z=k) &
G^{t^*}_{tk}(\tau) &\equiv \mathbb{P}(\varepsilon \leq \tau|T^*=t^*, T=t,z=k).
\end{align*}
Assumption \ref{assump:model} (i) imposes a relationship between $G^{t^*}_{tk}$ and $F^{t^*}_{tk}$ for each $t^*$, namely 
\begin{equation}
  G^0_{tk}(\tau) = F^0_{tk}(\tau + c), \quad
  G^1_{tk}(\tau) = F^1_{tk}(\tau + c + \beta)
  \label{eq:Gtstartk}
\end{equation}
and thus we see that
\begin{align}
  G_k(\tau) &= r_{1k}p_k F^1_{1k}(\tau + c + \beta) + r_{0k}(1 - p_k) F^1_{0k}(\tau + c + \beta) \nonumber \\
  &\quad +  (1 - r_{1k})p_k F^0_{1k}(\tau + c) + (1 - r_{0k})(1 - p_k) F^0_{0k}(\tau + c)
  \label{eq:Gk}
\end{align}
applying the law of total probability and Bayes' rule.
Moreover,
\begin{equation}
  F_{tk}(\tau) = r_{tk} F_{tk}^1(\tau) + (1 - r_{tk}) F_{tk}^0(\tau)
  \label{eq:Ftk}
\end{equation}
for all $t,k \in \left\{ 0,1 \right\}$, and by Bayes' rule,
\begin{equation}
  r_{1k} = (1 - \alpha_1)p_k^*/p_k, \quad
  r_{0k} = \alpha_1p_k^*/(1 - p_k).
  \label{eq:rtk}
\end{equation}
There are four cases, corresponding to different possibilities for the $r_{tk}$.
The first case violates one of our model assumptions.
For each of the remaining cases, we show that it is possible to construct the required distributions $F_{tk}^0, F_{tk}^1$ under Assumptions \ref{assump:model} (i) and (iii) for any $(c,\beta,\alpha_0, \alpha_1)$ such that $\mathbb{E}(y|z=k) = c + \beta(p_k - \alpha_0)/(1 - \alpha_0 - \alpha_1)$.

\paragraph{Case I: $r_{1k} = 0, r_{0k} \neq 0$}
By Equation \ref{eq:rtk} this requires $\alpha_1 = 1$, violating Assumption \ref{assump:misclassification} (ii).

\paragraph{Case II: $r_{0k} = r_{1k} = 0$}
By Equation \ref{eq:rtk}, this requires $p_k^* = 0$ which in turn requires $p_k = \alpha_0$.
By Equation \ref{eq:Ftk} we have $F^0_{tk} = F_{tk}$, while $F^1_{tk}$ is unrestricted.
Substituting into \ref{eq:Gk},
\[
  G_k(\tau) = p_k F_{1k}(\tau + c) + (1 - p_k) F_{0k}(\tau + c) = F_k(\tau + c)
\]
Now, since $F_k(\tau + c)$ is the conditional CDF of $y-c$ given that $z=k$, and $G_k$ is the conditional CDF of $\varepsilon$ given $z=k$,
we see that Assumption \ref{assump:model} (i) is satisfied if and only if $\mathbb{E}(y|z=k) = c$, which is equal to $c + \beta(p_k - \alpha_0) / (1 - \alpha_0 - \alpha_1)$ since $p_k - \alpha_0 = 0$.

\paragraph{Case III: $r_{1k}\neq 0, r_{0k} = 0$}
By Equation \ref{eq:rtk} this requires $\alpha_1 = 0$ and $p_k^* \neq 0$.
By Equation \ref{eq:Ftk} we have $F^0_{0k} = F_{0k}$ and since $r_{1k} \neq 1$, we can solve to obtain
\[
  F^1_{1k}(\tau) = \frac{1}{r_{1k}}\left[F_{1k}(\tau) - (1 - r_{1k})F^0_{1k}(\tau)\right]
\]
Substituting into Equation \ref{eq:Gk}, we obtain
\begin{align*}
  G_k(\tau) &= \left[ (1 - p_k)F_{0k}(\tau + c) + p_k F_{1k}(\tau + c + \beta) \right] \\ 
  &\quad + p_k(1 - r_{1k})\left[ F^0_{1k}(\tau + c) - F^0_{1k}(\tau + c + \beta) \right]
\end{align*}
Now, $F_{0k}(\tau + c)$ is the conditional CDF of $(y-c)$ given $(T=0,z=k)$ while $F_{1k}(\tau + c + \beta)$ is the conditional CDF of $(y-c-\beta)$ given $(T=1,z=k)$.
Similarly, $F^0_{1k}(\tau + c)$ is the conditional CDF of $\varepsilon$ given $(T^*=0,T = 1, z=k)$ while $F^0_{1k}(\tau + c + \beta)$ is the conditional CDF of $(\varepsilon - \beta)$ given $(T^*=0, T=1, z=k)$.
Since $G_k(\tau)$ is the conditional CDF of $\varepsilon$ given $z=k$, we see that Assumption \ref{assump:model} (iii) is satisfied if and only if
\begin{align*}
  0 &= (1 - p_k) \mathbb{E}(y-c|T=0,z=k) + p_k \mathbb{E}(y - c - \beta|T=1,z=k)\\
  &\quad + p_k(1 - r_{1k})\left[ \mathbb{E}(\varepsilon|T^*=0,T=1,z=k) - \mathbb{E}(\varepsilon - \beta|T^*=0,T=1,z=k) \right]
\end{align*}
Rearranging, this is equivalent to 
\[
  \mathbb{E}(y|z=k) = c + (1 - \alpha_1) \beta\left( \frac{p_k - \alpha_0}{1 - \alpha_0 - \alpha_1} \right) = c + \beta\left( \frac{p_k - \alpha_0}{1 - \alpha_0 - \alpha_1} \right)
\]
since $\alpha_1 = 0$ in this case.
As explained above, $F^0_{0k} = F_{0k}$ in the present case while $F^1_{0k}$ is undefined. 
We are free to choose any distributions for $F^{0}_{1k}$ and $F^{1}_{1k}$ that satisfy Equation \ref{eq:Ftk}, for example $F^{0}_{1k} = F^{1}_{1k} = F_{1k}$.

\paragraph{Case IV: $r_{1k}\neq 0, r_{0k} \neq 0$}
In this case, we can solve Equation \ref{eq:Ftk} to obtain
\[
  F^1_{tk}(\tau) = \frac{1}{r_{tk}}\left[F_{tk}(\tau) - (1 - r_{tk})F^0_{tk}(\tau)\right]
\]
Substituting this into Equation \ref{eq:Gk}, we have
\begin{align*}
  G_k(\tau) = F_k(\tau + c + \beta) &+ p_k(1 - r_{1k})\left[F^0_{1k}(\tau + c) - F_{1k}^0(\tau + c + \beta)\right]\\
  &+ (1 - p_k)(1 - r_{0k}) \left[ F^0_{0k}(\tau + c) - F^0_{0k}(\tau + c + \beta) \right]
\end{align*}
using the fact that $F_k(\tau) = p_k F_{1k}(\tau) + (1 - p_k) F_{0k}(\tau)$.
Now, $F_k(\tau + c + \beta)$ is the conditional CDF of $(y - c - \beta)$ given $z=k$, while $F_{tk}^0(\tau + c)$ is the conditional CDF of $\varepsilon$ given $(T = t,z =k)$ and $F^0_{tk}(\tau + c + \beta)$ is the conditional CDF of $(\varepsilon - \beta)$ given $(T = t, z=k)$.
Since $G_k(\tau)$ is the conditional CDF of $\varepsilon$ given $z=k$, we see that Assumption \ref{assump:model} (iii) is satisfied if and only if
\begin{align*}
  0 &= \mathbb{E}[y - c - \beta|z=k] + p_k(1 - r_{1k})\left[ \mathbb{E}(\varepsilon|T^*=0,T=1,z=k) - \mathbb{E}(\varepsilon - \beta|T^*=0,T=1,z=k) \right] \\
   &\quad + (1 - p_k)(1 - r_{0k})\left[ \mathbb{E}(\varepsilon|T^*=0,T=0,z=k) - \mathbb{E}(\varepsilon - \beta|T^*=0,T=0,z=k) \right]\\
   0 &= \mathbb{E}[y - c - \beta|z=k] + \beta\left[p_k(1 - r_{1k}) + (1 - p_k)(1 - r_{0k})\right]
\end{align*}
But since $\left[p_k(1 - r_{1k}) + (1 - p_k)(1 - r_{0k})\right] = (1 - p_k^*)$ and $p_k^* = (p_k - \alpha_0) /(1 - \alpha_0 - \alpha_1)$, this becomes 
\[
\mathbb{E}[y|z=k] = c + \beta\left[ (p_k - \alpha_0)(1 - \alpha_0 - \alpha_1) \right].
\]
Thus, in this case we are free to choose \emph{any} distributions for $F^{0}_{tk}$ and $F^1_{tk}$ that satisfy Equation \ref{eq:Ftk}.
For example we could take $F^0_{tk} = F^1_{tk} = F_{tk}$.
\end{proof}

\begin{proof}[Proof of Corollary \ref{cor:sharpBeta1}]
The result follows by substituting the largest and smallest possible values for $\alpha_0 + \alpha_1$ and taking the difference of the expressions for $\mathbb{E}[y|z=k]$.
\end{proof}


\begin{proof}[Proof of Theorem \ref{thm:sharpII}]
The only difference between the conditions of Theorem \ref{thm:sharpI} and those of \ref{thm:sharpII} is that the latter imposes Assumption \ref{assump:misclassification} (iii) while the former does not.
Accordingly, the present argument builds on the proof of Theorem \ref{thm:sharpI} and relies on the notation defined within it.
Under Assumption \ref{assump:model} (i), Assumption \ref{assump:misclassification} (iii) is equivalent to $\mathbb{E}[y|T,T^*,z] = \mathbb{E}[y|T^*,z]$.
Hence, non-differential measurement error constrains only the conditional distribution of $y$ given $(T,T^*,z)$.
For this reason, we need only revisit the second step of the proof of Theorem \ref{thm:sharpI}.
Consider a point $(c, \beta, \alpha_0, \alpha_1)$ that satisfies Equation \ref{eq:identsetI} and $\alpha_0 \leq p_k \leq 1 - \alpha_1$ for all $k$.
Since this point lies in the identified set from Theorem \ref{thm:sharpI}, it suffices to determine whether there exist valid conditional CDFs $F_{tk}^{0}, F_{tk}^{1}$ such that $F_{tk} = (1 - r_{tk}) F_{tk}^0 + r_{tk} F_{tk}^1$ for all $t,k$ and $\mathbb{E}[y|T,T^*,z] = \mathbb{E}[y|T^*,z]$.

Let $\mu_{tk}^{t^*} \equiv \mathbb{E}[y|T=t,z=k,T^*=t^*]$, $\mu_{tk} \equiv \mathbb{E}[y|T=t,z=k]$, and $\mu_{k}^{t^*} \equiv \mathbb{E}[y|z=k,T^*=t^*]$.
By Assumption \ref{assump:misclassification} (iii) $\mu_{tk}^{t^*} = \mu_{k}^{t^*}$ for $t^* =0,1$.
Hence, by iterated expectations,
\begin{align*}
  \mu_{0k} &= (1 - r_{0k}) \mu_{k}^0 + r_{0k} \mu_{k}^1\\ 
  \mu_{1k} &= (1 - r_{1k}) \mu_{k}^0 + r_{1k} \mu_{k}^1.
\end{align*}
Now, $(\mu_{0k},\mu_{1k})$ are observed while $r_{0k}$ and $r_{1k}$ depend only on the observed first-stage probability $p_k$ and the mis-classification probabilities $(\alpha_0,\alpha_1)$. Thus, at a given point $(c, \beta, \alpha_0, \alpha_1)$ in the identified set from Theorem \ref{thm:sharpI} the preceding equations form a linear system in $\mu_{k}^0$ and $\mu_{k}^1$.
After some algebra, we find that the determinant is
\[
  r_{1k} - r_{0k} = \left[ \frac{p_k - \alpha_0}{1 - \alpha_0 - \alpha_1} \right]\left[ \frac{1 - p_k - \alpha_1}{p_k(1 - p_k)} \right].
\]

Suppose first that $r_{0k} = r_{1k} = r$ so the determinant condition fails.
This occurs if and only if $\alpha_0 = p_k$ or $\alpha_1 = 1 - p_k$.
If $\mu_{0k} \neq \mu_{1k}$, the system is inconsistent: no solution for $(\mu_k^0, \mu_k^1)$ exists.
Hence $\alpha_0 = p_k$ and $\alpha_1 = 1 - p_k$ are excluded from the identified set under non-differential measurement error so long as $\mu_{0k} \neq \mu_{1k}$.
If instead $\mu_{0k} = \mu_{1k} = \mu$, the system is consistent but rank deficient: any pair $(\mu_k^0, \mu_k^1)$ such that $\mu = (1 - r) \mu_{k}^0 + r \mu_{k}^1$ is a solution and hence satisfies the assumption of non-differential measurement error.
One such solution is $\mu_{k}^1 = \mu_{k}^0 = \mu$ so we are free to set $F_{0k}^0 = F_{0k}^1 = F_{0k}$ and $F_{1k}^0 = F_{1k}^{1} = F_{1k}$.
Hence, if $\mu_{0k} = \mu_{1k}$ then $\alpha_0 = p_k$ lies within the sharp identified set if $p_k < p_\ell$ and $\alpha_1 = 1 - p_k$ lies in the sharp identified set if $p_\ell < p_k$.

Now suppose that $r_{0k} \neq r_{1k}$, which occurs if and only if $\alpha_0 \neq p_k$ and $\alpha_1 \neq 1 - p_k$.
In this case the system has a unique solution, namely
\begin{align*}
  \mu_k^0 &= \frac{r_{1k}\mu_{0k} - r_{0k}\mu_{1k}}{r_{1k} - r_{0k}} = \frac{(1 - p_k) \mathbb{E}(y|T=0,z=k) - \alpha_1 \mathbb{E}(y|z=k)}{1 - p_k - \alpha_1} \\ 
  \mu_k^1 &= \frac{(\mu_{1k} - \mu_{0k}) + (r_{1k}\mu_{0k} - r_{0k}\mu_{1k})}{r_{1k} - r_{0k}} = \frac{p_k \mathbb{E}(y|T=1,z=k) - \alpha_0 \mathbb{E}(y|z=k)}{p_k - \alpha_0}.
\end{align*}
Since $\mu_{k}^0 = \mu_{0k}^0 = \mu_{1k}^0$ and $\mu_{k}^1 = \mu_{0k}^1 = \mu_{1k}^1$ under non-differential measurement error, the mis-classification probabilities $(\alpha_0, \alpha_1)$ combined with the observable moments completely determine the means of $F_{tk}^0$ and $F_{tk}^1$ whenever the determinant condition holds.  
If $\mu_{0k} = \mu_{1k}$ then $\mu_{k}^0 = \mu_k^1$ so we are free to set $F_{0k}^0 = F_{0k}^1 = F_{0k}$ and $F_{1k}^0 = F_{1k}^1 = F_{1k}$.
Combining this with the reasoning from the preceding paragraph, we see that Assumption \ref{assump:misclassification} (iii) imposes \emph{no additional restrictions} for any $k$ such that $\mu_{0k} = \mu_{1k}$.
Accordingly, for the remainder of the proof we consider only the case in which $\mu_{0k} \neq \mu_{1k}$.
Given $(\alpha_0, \alpha_1)$, $r_{tk}, \mu^0_k$, and $\mu^1_{k}$ are fixed.
The question is whether, for a given pair $(\alpha_0, \alpha_1)$ and observed CDFs $F_{tk}$, we can construct valid CDFs $F_{tk}^0, F_{tk}^1$ such that 
\[
  \int_{\mathbb{R}} \tau F_{tk}^0(d\tau) = \mu_k^0, \quad
  \int_{\mathbb{R}} \tau F_{tk}^1(d\tau) = \mu_k^1, \quad 
  F_{tk}(\tau) = r_{tk} F^1_{tk}(\tau) + (1 - r_{tk}) F^0_{tk}(\tau).
\]
For a given pair $(t,k)$, there are two cases: $0 < r_{tk} < 1$ and $r_{tk} \in \left\{ 0, 1 \right\}$.

\paragraph{Case I: $r_{tk}\in \left\{ 0,1 \right\}$}
If $r_{tk} = 1$ then $\mu^1_k = \mu_{tk}$ so we can set $F^1_{tk} = F_{tk}$.
In this case $F^0_{tk}$ is unrestricted.
Analogously, if $r_{tk} = 0$, $\mu^0_k = \mu_{tk}$ so we can set $F^0_{tk} = F_{tk}$ with $F^1_{tk}$ unrestricted.

\paragraph{Case II: $0 < r_{tk} < 1$} 
Define the function $\mu_{tk}(\xi) = \mathbb{E}[y|y\in I_{tk}(\xi), T=t, z=k]$ and the closed interval $I_{tk}(\xi) = \left[ F^{-1}_{tk}(1 - \xi - r_{tk}), F^{-1}_{tk}(1 - \xi) \right]$ where $0 \leq \xi \leq 1 - r_{tk}$.
The function $\mu_{tk}$ is decreasing in $\xi$,  attaining its maximum $\overline{\mu}_{tk}$ at $\xi = 0$ and its minimum $\underline{\mu}_{tk}$ at $\xi = 1 - r_{tk}$.

Suppose first that $\mu^1_{k}$ does \emph{not} lie in the interval $[\underline{\mu}_{tk}, \overline{\mu}_{tk}]$.
We show that it is impossible to construct valid CDFs $F^0_{tk}$ and $F^{1}_{tk}$ that satisfy $F_{tk}(\tau) = r_{tk} F^1_{tk}(\tau) + (1 - r_{tk}) F^0_{tk}(\tau)$.
Since $r_{tk} \neq 1$, we can solve the expression for $F_{tk}$ to yield 
  $F^{0}_{tk}(\tau) = \left[ F_{tk}(\tau) - r_{tk} F^1_{tk}(\tau)\right] / (1 - r_{tk})$.
  Hence, since $r_{tk} \neq 0$, the requirement that $0 \leq F_{tk}^0(\tau) \leq 1$ implies
\begin{equation}
  \frac{F_{tk}(\tau) - (1 - r_{tk})}{r_{tk}} \leq F^{1}_{tk}(\tau) \leq \frac{F_{tk}(\tau)}{r_{tk}}
  \label{eq:F1tk_ineq}
\end{equation}
Now define $\underline{F}^{1}_{tk}(\tau) =  \min\left\{ 1,\,  F_{tk}(\tau)/r_{tk} \right\}$ and 
$\overline{F}^{1}_{tk}(\tau) = \max\left\{ 0,\,  F_{tk}(\tau)/r_{tk} - (1 - r_{tk})/r_{tk} \right\}$.
By combining Equation \ref{eq:F1tk_ineq} with $0 \leq F^{1}_{tk}(\tau) \leq 1$, we obtain $\overline{F}_{tk}^1(\tau) \leq F^{1}_{tk}(\tau) \leq \underline{F}_{tk}^1(\tau)$.
Thus, $\overline{F}^1_{tk}$ first-order stochastically dominates $F^{1}_{tk}$ which first-order stochastically dominates $\underline{F}_{tk}^1$. 
Hence,
\[
 \int \tau \underline{F}_{tk}^1(d\tau) \leq \int \tau F^{1}_{tk}(d\tau) \leq \int \tau\overline{F}_{tk}^1(d\tau).
\]
But notice that 
\[
  \underline{\mu}_{tk} = \int \tau \underline{F}_{tk}^1(d\tau), \quad 
  \mu^1_{k} = \int \tau F^{1}_{tk}(d\tau), \quad 
  \overline{\mu}_{tk} = \int \tau\overline{F}_{tk}^1(d\tau)
\]
so we have $\underline{\mu}_{tk} \leq \mu^1_{k} \leq \overline{\mu}_{tk}$ which contradicts $\mu^1_{k} \notin [\underline{\mu}_{tk}, \overline{\mu}_{tk}]$.

Now suppose that $\mu^1_{k} \in \left[\underline{\mu}_{tk}, \overline{\mu}_{tk} \right]$.
We show how to construct densities $f_{tk}^1$ and $f_{tk}^0$ that yield CDFs $F_{tk}^0$ $F_{tk}^1$ satisfying the requirements described above.
Since the conditional distribution of $y$ given $(T,z)$ is continuous, $\mu_{tk}$ is continuous on its domain and takes on all values in $\left[ \underline{\mu}_{tk}, \overline{\mu}_{tk} \right]$ by the intermediate value theorem.
Thus, there exists a $\xi^*$ such that $\mu_{tk}(\xi^*) = \mu^1_{k}$.
Let $f_{tk}(\tau) = dF_{tk}(\tau)/d\tau$ which is non-negative by the assumption that $y$ is continuously distributed.
Now, define
\[
  f^1_{tk}(\tau) = \frac{f_{tk}(\tau)\times \mathbf{1}\left\{ \tau \in I_{tk}(\xi^*) \right\}}{r_{tk}}, \quad
  f^0_{tk}(\tau) = \frac{f_{tk}(\tau) \times \mathbf{1}\left\{ \tau \in I_{tk}(\xi^*) \right\}}{1 - r_{tk}}.
\]
Clearly $f_{tk}^1\geq 0$ and $f^0_{tk} \geq 0$.
Integrating, 
\begin{align*}
  \int_{\mathbb{R}} f_{tk}^1(\tau) \; d\tau = \frac{1}{r_{tk}}\int_{I_{tk}(\xi^*)} f_{tk}(\tau)\; d\tau = 1, \quad
  \int_{\mathbb{R}} f_{tk}^0(\tau) \; d\tau = \frac{1}{1 - r_{tk}}\int_{I^C_{tk}(\xi^*)} f_{tk}(\tau)\; d\tau = 1
\end{align*}
where $I_{tk}^C$ is the complement of $I_{tk}$.
By construction
\[
  r_{tk} \int_A f_{tk}^1(\tau) \; d\tau + (1 - r_{tk}) \int_A f_{tk}^0(\tau) \; d\tau = \int_A f_{tk}(\tau)\; d\tau
\]
for any set $A$. 
Finally,
\[
  \int_{\mathbb{R}} \tau f_{tk}^1(\tau) \; d\tau = \frac{1}{r_{tk}} \int_{I_{tk}(\xi^*)} \tau f_{tk}(\tau)\; d\tau = \mu_{tk}(\xi^*) = \mu^1_{k}.
\]
\end{proof}

\subsection{Point Identification Results}
In the proofs of Lemma \ref{lem:eta2}, Lemma \ref{lem:eta3}, and Theorem \ref{thm:main_ident}, we employ the shorthand $\pi \equiv \mbox{Cov}(T,z)$, $\eta_j \equiv \mbox{Cov}(y^j,z)$, and $\tau_j \equiv \mbox{Cov}(Ty^j,z)$ for $j = 1, 2, 3$.
Hence Lemma \ref{lem:wald} becomes $\eta_1 = \pi\theta_1$, while Lemma \ref{lem:eta2} becomes $\eta_2 =  2\tau_1 \theta_1 - \pi \theta_2$, 
and Lemma \ref{lem:eta3} becomes $\eta_3 = 3\tau_2 \theta_1 - 3\tau_1 \theta_2 + \pi\theta_3$.

\begin{proof}[Proof of Lemma \ref{lem:eta2}]
  By Assumption \ref{assump:model} (i) and the basic properties of covariance, 
\begin{align*}
    \eta_2 &= \beta^2 \mbox{Cov}(T^*,z) + 2 \beta\left[ c\, \mbox{Cov}(T^*,z) + \mbox{Cov}(T^*\varepsilon,z)  \right] + 2c \, \mbox{Cov}(\varepsilon,z) + \mbox{Cov}(\varepsilon^2,z)\\
  \tau_1 &= c \pi + \mbox{Cov}(T\varepsilon,z) + \beta \mbox{Cov}(TT^*,z)
\end{align*}
using the fact that $T^*$ is binary. 
Now, by Assumptions \ref{assump:model} (iii) and \ref{assump:2ndMoment} we have $\mbox{Cov}(\varepsilon,z) = \mbox{Cov}(\varepsilon^2,z) = 0$.
And, using Assumptions \ref{assump:misclassification} (i) and (ii), one can show that $\mbox{Cov}(TT^*,z) = (1 - \alpha_1)\mbox{Cov}(T^*,z)$ and $\mbox{Cov}(T^*,z) = \pi/(1 - \alpha_0 - \alpha_1)$.
Hence, 
\begin{align*}
  \eta_2 &= \theta_1\left( \beta 
+ 2 c \right) \pi + 2\beta \mbox{Cov}(T^*\varepsilon,z) \\
  2 \tau_1 \theta_1 - \pi \theta_2 &= \left[2\theta_1 c + 2 \theta_1^2 (1 - \alpha_1) - \theta_2\right]\pi + 2\theta_1 \mbox{Cov}(T\varepsilon,z) 
\end{align*}
but since $\theta_2 = \theta_1^2 \left[ (1 - \alpha_1) + \alpha_0 \right]$, we see that $[2\theta_1^2(1 - \alpha_1) - \theta_2] = \theta_1 \beta$.
Thus, it suffices to show that $\beta \mbox{Cov}(T^*\varepsilon,z) = \theta_1 \mbox{Cov}(T\varepsilon,z)$.
This equality is trivially satisfied when $\beta=0$, so suppose that $\beta \neq 0$. 
In this case it suffices to show that $(1 - \alpha_0 - \alpha_1) \mbox{Cov}(T^*\varepsilon,z) = \mbox{Cov}(T\varepsilon,z)$.
Define $m^*_{tk} = \mathbb{E}\left[ \varepsilon|T^*=t,z=k \right]$ and $p^*_k = \mathbb{P}(T^*=1|z=k)$.
Then, by iterated expectations, Bayes' rule, and Assumption \ref{assump:misclassification} (iii)
\begin{align*}
  \mbox{Cov}(T^*\varepsilon,z) &=q(1 - q)\left(p_1^* m_{11}^*  - p_0^*m_{10}^* \right) \\
  \mbox{Cov}(T\varepsilon,z) &= q(1 - q)\left\{ (1 - \alpha_1)\left[ p_1^* m_{11}^* - p_0^* m_{10}^* \right] + \alpha_0\left[ (1 - p_1^*) m_{01}^* - (1 - p_0^*)m_{00}^* \right] \right\} 
\end{align*}
But by Assumption \ref{assump:model} (iii), $\mathbb{E}[\varepsilon|z=k] = m_{1k}^*p_{k}^* + m_{0k}^*(1 - p_k^*)=0$ and thus we obtain $m_{0k}^*(1 - p_k^*)= - m_{1k}^* p_k^*$.
Therefore  $(1 - \alpha_0 - \alpha_1) \mbox{Cov}(T^*\varepsilon,z) = \mbox{Cov}(T\varepsilon,z)$ as required.
\end{proof}

\begin{proof}[Proof of Lemma \ref{lem:eta3}]
  Since $T^*$ is binary, if follows from the basic properties of covariance that,
\begin{align*}
  \eta_3 &= \mbox{Cov}\left[ (c + \varepsilon)^3,z \right] + 3 \beta \mbox{Cov}[(c + \varepsilon)^2 T^*, z] + 3 \beta^2 \mbox{Cov}[(c + \varepsilon)T^*,z] + \beta^3 \mbox{Cov}(T^*,z)\\
  \tau_2 &= \mbox{Cov}\left[ (c + \varepsilon)^2 T, z \right] + 2 \beta \mbox{Cov}\left[ (c + \varepsilon)TT^*,z \right] + \beta^2 \mbox{Cov}(TT^*,z)
\end{align*}
By Assumptions \ref{assump:model} (iii), \ref{assump:2ndMoment}, and \ref{assump:3rdMoment} (ii) , $\mbox{Cov}\left[ (c + \varepsilon)^3,z \right] = 0$.
Expanding, 
\begin{align*}
  \eta_3 
  &= 3 \beta \mbox{Cov}(T^*\varepsilon^2,z) + \left(3 \beta^2 + 6c\beta \right)\mbox{Cov}(T^*\varepsilon,z) + \left( \beta^3 + 3c\beta^2 + 3c^2\beta \right)\mbox{Cov}(T^*, z)\\
  \tau_2 &= c^2 \mbox{Cov}(T,z) + \beta(\beta + 2c) \mbox{Cov}(TT^*,z) + \mbox{Cov}(T\varepsilon^2,z) + 2c \mbox{Cov}(T\varepsilon,z) + 2\beta\,\mbox{Cov}(TT^*\varepsilon,z)
\end{align*}
Now, define $s^*_{tk} = \mathbb{E}[\varepsilon^2|T^*=t, z=k]$ and $p_k^* = \mathbb{P}(T^*=1|z=k)$.
By iterated expectations, Bayes' rule, and Assumption \ref{assump:3rdMoment} (i), 
\begin{align*}
  \mbox{Cov}(T^*\varepsilon^2, z) 
  &= q(1 - q)(p^*_1 s^*_{11} - p^*_0 s^*_{10}) \\
  \mbox{Cov}(T\varepsilon^2, z) 
  &= q(1 - q)\left\{ (1 - \alpha_1)\left[p^*_1 s_{11}^* - p_0^* s_{10}^*\right] + \alpha_0 \left[ (1 - p_1^*)s_{01}^* - (1 - p_0^*) s_{00}^*\right] \right\}
\end{align*}
By Assumption \ref{assump:2ndMoment}, $\mathbb{E}[\varepsilon^2|z=1] = \mathbb{E}[\varepsilon^2|z=0]$ and thus, by iterated expectations we have
$p_1^* s_{11}^* - p_0^* s^*_{10} =  - \left[(1 - p_1^*)s_{01}^* - (1 - p_0^*)s_{00}^* \right]$
which implies 
\begin{equation}
  \mbox{Cov}(T\varepsilon^2,z) = (1 - \alpha_0 - \alpha_1)\mbox{Cov}(T^*\varepsilon^2,z).
  \label{eq:TEpsilonSquared}
\end{equation}
Similarly by iterated expectations and Assumptions \ref{assump:misclassification} (i)--(ii)
\begin{equation}
  \mbox{Cov}(TT^*\varepsilon, z) = q(1 - q)(1 - \alpha_1)(p_1^* m_{1k}^* - p_0^* m_{10}^*) = (1 - \alpha_1) \mbox{Cov}(T^*\varepsilon, z) 
  \label{eq:TTstarEpsilon}
\end{equation}
where $m_{tk}^*$ is defined as in the proof of Lemma \ref{lem:eta2}.
As shown in the proof of Lemma \ref{lem:eta2}, 
\begin{align*}
  \mbox{Cov}(TT^*,z) = (1 - \alpha_1) \mbox{Cov}(T^*,z), \quad
  \mbox{Cov}(T^*,z) = \frac{\pi}{1 - \alpha_0 - \alpha_1}, \quad
  \mbox{Cov}(T^*\varepsilon,z) = \frac{\mbox{Cov}(T\varepsilon,z)}{1 - \alpha_0 - \alpha_1}
\end{align*}
and combining these equalities with Equations \ref{eq:TEpsilonSquared} and \ref{eq:TTstarEpsilon}, it follows that
\begin{align*}
  \tau_2 &=  2\left[(1 - \alpha_1)(c + \beta) - c \alpha_0\right]\mbox{Cov}(T^*\varepsilon,z) + \left[(1 - \alpha_1)(c + \beta)^2 - c^2 \alpha_0 \right]\mbox{Cov}(T^*,z)\\
  &\quad \quad +(1 - \alpha_0 - \alpha_1)\mbox{Cov}(T^*\varepsilon^2,z) \\
  \tau_1 &= (1 - \alpha_0 - \alpha_1)\mbox{Cov}(T^*\varepsilon,z) + \left[(1 - \alpha_1)(c + \beta) - c \alpha_0\right] \mbox{Cov}(T^*,z)
\end{align*}
using $\tau_1 = c\pi + \mbox{Cov}(T\varepsilon,z) + \beta\mbox{Cov}(TT^*,z)$ as shown in the proof of Lemma \ref{lem:eta2}.
Thus, 
\[
  3\tau_2 \theta_1 - 3 \tau_1 \theta_2 + \pi \theta_3 = K_1 \mbox{Cov}(T^*\varepsilon^2,z) + K_2 \mbox{Cov}(T^*\varepsilon, z) + K_3 \mbox{Cov}(T^*,z)
\]
where $K_1 \equiv 3 \theta_1(1 - \alpha_0 - \alpha_1) = 3 \beta$ and
\begin{align*}
  K_2 &\equiv 6\theta_1 \left[(1 - \alpha_1)(c +\beta) - c\alpha_0\right] - 3\theta_2 (1 - \alpha_0 - \alpha_1) \\
  K_3 &\equiv 3\theta_1 \left[ (1 - \alpha_1)(c + \beta)^2 - c^2 \alpha_0 \right] - 3\theta_2 \left[ (1 - \alpha_1)(c + \beta) - c\alpha_0 \right] + \theta_3(1 - \alpha_0 - \alpha_1)
\end{align*}
Substituting the definitions of $\theta_1, \theta_2$, and $\theta_3$ from Equations \ref{eq:theta1_def}--\ref{eq:theta3_def}, tedious but straightforward algebra shows that $K_2 = 3\beta^2 + 6c\beta$ and $K_3 = \beta^3 + 3c\beta^2 + 3c^2\beta$.
Therefore the coefficients of $\eta_3$ equal those of $3\tau_2 - 3\tau_1 \theta_2 + \pi \theta_3$ and the result follows.
\end{proof}


\begin{proof}[Proof of Theorem \ref{thm:main_ident}]
  Collecting the results of Lemmas \ref{lem:wald}--\ref{lem:eta3}, we have
\[
 \eta_1 = \pi\theta_1, \quad
  \eta_2 =  2\tau_1 \theta_1 - \pi \theta_2, \quad
  \eta_3 = 3\tau_2 \theta_1 - 3\tau_1 \theta_2 + \pi\theta_3
\]
which is a linear system in $\theta_1, \theta_2, \theta_3$ with determinant $-\pi^3$.
Since $\pi \neq 0$ by assumption \ref{assump:model} (ii), $\theta_1, \theta_2$ and $\theta_3$ are identified.
  Now, so long as $\beta \neq 0$, we can rearrange Equations \ref{eq:theta2_def} and \ref{eq:theta3_def} to obtain 
  \begin{align}
    \label{eq:quadraticA}
  A &= \theta_2/\theta_1^2 = 1 + (\alpha_0 - \alpha_1)  \\
  \label{eq:quadraticB}
  B &= \theta_3/\theta_1^3 = (1 - \alpha_0 - \alpha_1)^2 + 6 \alpha_0 (1 - \alpha_1)
  \end{align}
  Equation \ref{eq:quadraticA} gives $(1 - \alpha_1)= A - \alpha_0$.
  Hence $(1 - \alpha_0 - \alpha_1) = A - 2\alpha_0$ and $\alpha_0(1 - \alpha_1) = \alpha_0(A - \alpha_0)$.
  Substituting into Equation \ref{eq:quadraticB} and simplifying, $(A^2 - B) + 2A \alpha_0 - 2\alpha_0^2=0$.
  Substituting for $\alpha_0$ analogously yields a quadratic in $(1 - \alpha_1)$ with \emph{identical} coefficients.
It follows that one root of $(A^2-B) + 2Ar - 2r^2=0$ is $\alpha_0$ and the other is $1 - \alpha_1$.
Solving,
  \begin{equation}
    r = \frac{A}{2} \pm \sqrt{3 A^2 - 2B} = \frac{1}{\theta_1^2}\left(\frac{\theta_2}{2} \pm  \sqrt{3\theta_2^2  - 2\theta_1 \theta_3}\right).
  \end{equation}
Substituting Equations \ref{eq:theta2_def} and \ref{eq:theta3_def}, simple algebra shows that $3\theta_2^2 - 2 \theta_1\theta_3 = \theta_1^2(1 - \alpha_0 - \alpha_1)^2$.
This quantity is strictly greater than zero since $\theta_1 \neq 0$ and $\alpha_0 + \alpha_1 \neq 1$.
It follows that both roots of the quadratic are real.
Moreover, $3\theta_2^2/\theta_1^4 - 2\theta_3/\theta_1^3$ identifies $(1 - \alpha_0 - \alpha_1)^2$.
Substituting into Equation \ref{eq:theta1_def}, it follows that $\beta$ is identified up to sign.
If $\alpha_0 + \alpha_1 < 1$ then $\mbox{sign}(\beta) = \mbox{sign}(\theta_1)$ so that both the sign and magnitude of $\beta$ are identified.
If $\alpha_0 + \alpha_1 < 1$ then $1 - \alpha_1 > \alpha_0$ so $(1 - \alpha_1)$ is the larger root of $(A^2 - B) + 2Ar - 2r^2=0$ and $\alpha_0$ is the smaller root.
\end{proof}

\section{Comment on \cite{Mahajan} A.2}
\label{sec:mahajan}
Expanding on our discussion from Section \ref{sec:ident_literature} above, we now show that \citeauthor{Mahajan}'s identification argument for an endogenous regressor in an additively separable model (A.2) is incorrect.
Unless otherwise indicated, all notation used below is as defined in Section \ref{sec:identification}.

The first step of \cite{Mahajan} A.2 argues (correctly) that under Assumptions \ref{assump:model} and \ref{assump:misclassification} (i)--(ii), knowledge of $\alpha_0(\mathbf{x})$ and $\alpha_1(\mathbf{x})$ is sufficient to identify $\beta(\mathbf{x})$. 
This step is equivalent to our Lemma \ref{lem:wald} above.
The second step appeals to \cite{Mahajan} Theorem 1 to argue that $\alpha_0(\mathbf{x})$ and $\alpha_1(\mathbf{x})$ are indeed point identified.
To understand the logic of this second step, we first re-state \cite{Mahajan} Theorem 1 in our notation.
As in Section \ref{sec:identification} above, $T^*$ denotes an unobserved binary random variable, $z$ is a instrument, $T$ an observed binary surrogate for $T^*$, $y$ an outcome of interest, and $\mathbf{x}$ a vector covariates.

\begin{assump}[\cite{Mahajan} Theorem 1]
  Define $g(T^*, \mathbf{x}) \equiv \mathbb{E}[y|\mathbf{x},T^*]$ and $v \equiv y - g(T^*,\mathbf{x})$.
  Suppose that knowledge of $(y,T^*,\mathbf{x})$ is sufficient to identify $g$ and that:
  \begin{enumerate}[(i)]
    \item $\mathbb{P}(T^*=1|\mathbf{x},z=0) \neq \mathbb{P}(T^*=1|\mathbf{x},z=1)$.
    \item $T$ is conditionally independent of $z$ given $(\mathbf{x}, T^*)$.
    \item $\alpha_0(\mathbf{x}) + \alpha_1(\mathbf{x}) < 1$
    \item $\mathbb{E}[v|\mathbf{x},z,T^*,T] = 0$
    \item $g(1,\mathbf{x}) \neq g(0, \mathbf{x})$
  \end{enumerate}
  \label{assump:mahajan1}
\end{assump}

\begin{thm}[\cite{Mahajan} Theorem 1]
  Under Assumption \ref{assump:mahajan1}, $\alpha_0(\mathbf{x})$ and $\alpha_1(\mathbf{x})$ are point identified, as is $g(T^*,\mathbf{x})$.
  \label{thm:mahajan1}
\end{thm}

Assumption \ref{assump:mahajan1} (i) is equivalent to our Assumption \ref{assump:model} (ii), while Assumptions \ref{assump:mahajan1} (ii)--(iii) are equivalent to our Assumptions \ref{assump:misclassification} (i)--(ii).
Assumption \ref{assump:mahajan1} (v) serves the same purpose as $\beta(\mathbf{x}) \neq 0$ in our Theorem \ref{thm:main_ident}: unless $T^*$ affects $y$, we cannot identify the mis-classification probabilities.
The key difference between Theorem \ref{thm:mahajan1} and the setting we consider in Section \ref{sec:identification} comes from Assumption \ref{assump:mahajan1} (iv). 
This is essentially a stronger version of our Assumptions \ref{assump:model} (iii) and \ref{assump:misclassification} (iii) but applies to the \emph{projection error} $v$, defined in Assumption \ref{assump:mahajan1} rather than the structural error $\varepsilon$, defined in Assumption \ref{assump:model} (i).
Accordingly, Theorem \ref{thm:mahajan1} identifies the conditional mean function $g$ rather than the causal effect $\beta(\mathbf{x})$.

Although the meaning of the error term changes when we move from a structural to a reduced form model, the meaning of the mis-classification error rates does not: $\alpha_0(\mathbf{x})$ and $\alpha_1(\mathbf{x})$ are simply conditional probabilities for $T$ given $(T^*,\mathbf{x})$.
Step 2 of \cite{Mahajan} A.2 relies on this insight.
The idea is to find a way to satisfy Assumption \ref{assump:mahajan1} (iv) simultaneously with Assumptions \ref{assump:model} (iii) and \ref{assump:misclassification} (iii), while allowing $T^*$ to be endogenous.
If this can be achieved, $\alpha_0(\mathbf{x}), \alpha_1(\mathbf{x})$ will be identified via Theorem \ref{thm:mahajan1}, and identification of $\beta(\mathbf{x})$ will follow from step 1 of A.2 (our Lemma \ref{lem:wald}).
To this end, \cite{Mahajan} invokes the condition 
\begin{equation}
  \mathbb{E}(y|\mathbf{x},z,T^*,T) = \mathbb{E}(y|\mathbf{x},T^*).
  \label{eq:mahajan11}
\end{equation}
Because \cite{Mahajan} A.2 assumes an additively separable model -- our Assumption \ref{assump:model} (i) -- we see that
\[
  \mathbb{E}(y|\mathbf{x},z,T^*,T) = c(\mathbf{x}) + \beta(\mathbf{x}) T^* + \mathbb{E}(\varepsilon|\mathbf{x},z,T^*,T)
\]
so Equation \ref{eq:mahajan11} is equivalent to $\mathbb{E}(\varepsilon|\mathbf{x},z,T^*,T)=\mathbb{E}(\varepsilon|\mathbf{x},T^*)$.
Note that this allows $T^*$ to be endogenous, as it does not require $\mathbb{E}(\varepsilon|\mathbf{x},T^*)=0$.
Now, applying Equation \ref{eq:mahajan11} to the definition of $v$ from Assumption \ref{assump:mahajan1}, we have
\[
  \mathbb{E}(v|\mathbf{x},z,T^*,T) = \mathbb{E}\left[ y - \mathbb{E}(y|\mathbf{x},T^*)\left. \right|\mathbf{x},z,T^*,T \right] = 0
\]
which satisfies Assumption \ref{assump:mahajan1} (iv) as required.
Based on this reasoning, \cite{Mahajan} claims that Equation \ref{eq:mahajan11} along with Assumptions \ref{assump:mahajan1} (iv), \ref{assump:model}, and \ref{assump:misclassification} (i)--(ii) suffice to identify the effect $\beta(\mathbf{x})$ of an endogenous $T^*$, so long as $g(1,\mathbf{x}) \neq g(0,\mathbf{x})$.
As we now show, however, these Assumptions are contradictory unless $T^*$ is exogenous.

By Equation \ref{eq:mahajan11} and Assumption \ref{assump:model} (i), $\mathbb{E}(\varepsilon|\mathbf{x},z,T^*,T)=\mathbb{E}(\varepsilon|\mathbf{x},T^*)$ and thus by iterated expectations, we obtain
\begin{equation}
  \mathbb{E}(\varepsilon|\mathbf{x},T^*,z) = \mathbb{E}_{T|\mathbf{x},T^*,z}\left[ \mathbb{E}(\varepsilon|\mathbf{x},T^*,T,z) \right] = \mathbb{E}_{T|\mathbf{x},T^*,z}\left[ \mathbb{E}(\varepsilon|\mathbf{x},T^*) \right] = \mathbb{E}(\varepsilon|\mathbf{x}, T^*).
  \label{eq:zdiffs}
\end{equation}
Now, let $m^*_{tk}(\mathbf{x}) = \mathbb{E}(\varepsilon|\mathbf{x}, T^*=t,z=k)$.
Using this notation, Equation \ref{eq:zdiffs} is equivalent to $m^*_{t0}(\mathbf{x}) = m^*_{t1}(\mathbf{x})$ for $t = 0, 1$.
Combining iterated expectations with Assumption \ref{assump:model} (iii), 
\begin{equation}
  \mathbb{E}(\varepsilon|\mathbf{x},z=k) = [1 - p^*_k(\mathbf{x})] m^*_{0k}(\mathbf{x}) + p^*_k(\mathbf{x}) m^*_{1k}(\mathbf{x}) = 0 
  \label{eq:eps}
\end{equation}
for $k = 0,1$ where $p^*_k(\mathbf{x}) \equiv \mathbb{P}(T^*=1|\mathbf{x}, z=k)$.
But substituting $m^*_{t0}(\mathbf{x}) = m^*_{t1}(\mathbf{x})$ into Equation \ref{eq:eps} for $k=0,1$, we obtain 
\begin{align*}
  [1 - p^*_0(\mathbf{x})] m^*_{00}(\mathbf{x}) + p^*_0(\mathbf{x}) m^*_{10}(\mathbf{x}) &= 0\\ 
  [1 - p^*_1(\mathbf{x})] m^*_{00}(\mathbf{x}) + p^*_1(\mathbf{x}) m^*_{10}(\mathbf{x}) &= 0
\end{align*}
The preceding two equalities are convex combinations of $m^*_{00}$ and $m^*_{10}$.
The only way that both can equal zero simultaneously is if either $p^*_0(\mathbf{x}) = p^*_1(\mathbf{x})$, contradicting Assumption \ref{assump:model} (ii), or if $m^*_{tk}(\mathbf{x}) = 0$ for all $(t,k)$, which implies that $T^*$ is exogenous.
Hence \cite{Mahajan} A.2 fails: given the assumption that $z$ is a valid instrument for $\varepsilon$, Equation \ref{eq:mahajan11} implies that either there is no first-stage relationship between $z$ and $T^*$ or that $T^*$ is exogenous.
The root of the problem with A.2 is the attempt to use \emph{one} instrument to satisfy both the assumptions of Theorem \ref{thm:mahajan1} and Lemma \ref{lem:wald}.
If one had access to a second instrument $w$, or equivalently a second mis-measured surrogate for $T^*$, that satisfied Assumptions \ref{assump:mahajan1}, one could use $w$  to recover $\alpha_0(\mathbf{x})$ and $\alpha_1(\mathbf{x})$ via Theorem \ref{thm:mahajan1} and $z$ to recover the IV estimand $\beta(\mathbf{x}) / [1 - \alpha_0(\mathbf{x}) - \alpha_1(\mathbf{x})]$ via Lemma \ref{lem:wald}.
\section{Unobserved Heterogeneity}
\label{sec:het}
While allowing for arbitrary observed heterogeneity through the covariates $\mathbf{x}$, all of the results presented above assume an additively separable model -- Assumption \ref{assump:model} (i).
In this section we briefly discuss how our partial identification results can be interpreted in a local average treatment effects (LATE) setting.
For simplicity, we suppress explicit conditioning on the covariates $\mathbf{x}$ throughout. 

In lieu of Assumption \ref{assump:model} (i), consider a non-separable model of the form $y = h(T^*,z,\varepsilon)$.
Let $T^*(z)$ denote an individual's potential treatment and $Y(t^*,z)$ denote her potential outcome, where $t^*,z\in \left\{ 0,1 \right\}$.
Using this notation we can write $Y(t^*,z) = h(t^*,z,\varepsilon)$.
Let $J \in \left\{ a, c, d, n \right\}$ index the four LATE principal strata: $a = $ always-taker, $c = $ complier, $d = $ defier, and $n =$ never-taker.
If $J=a$, then $T^*(z) = 1$; if $J=c$, then $T^*(z) = z$; if $J=d$, then $T^*(z) = 1 - z$; and if $J=n$, then $T^*(z)=0$.
In a LATE model, Assumption \ref{assump:model} (iii) is replaced by the standard LATE assumptions:
\begin{assump}[Unconfounded Type]
  \label{assump:LATEtype}
    $\mathbb{P}(J=j|z=1) = \mathbb{P}(J=j|z=0)$ 
  for all  $j\in \left\{ a, c, d, n \right\}$.
\end{assump}
\begin{assump}[Mean Exclusion Restriction]
  \label{assump:LATEexclude}
    For all $t^* \in \left\{ 0,1 \right\}$ and $j\in \left\{ a, c, d, n \right\}$,
    \[\mathbb{E}\left[Y(t^*,0)|T^*=t^*,z=1\right]=\mathbb{E}\left[Y(t^*,1)|T^*=t^*,z=1\right] = \mathbb{E}[Y(t^*)|J=j].\]
\end{assump}
\begin{assump}[Monotonicity]
  \label{assump:LATEmono}
    $\mathbb{P}\big(T^*(1) \geq T^*(0)\big) = 1$ 
\end{assump}

As is well known, Assumption \ref{assump:model} (iii) combined with the preceding three conditions implies that the  instrumental variables estimand based on $T^*$ identifies the average treatment effect among compliers:
\[
  \frac{\mathbb{E}[y|z=1] - \mathbb{E}[y|z=0]}{p^*_1 - p^*_0} = \mathbb{E}[Y(1) - Y(0)|J=c].
\]
The numerator of the preceding expression is observed, but under mis-classification the denominator is not.
Notice, however, that Assumptions \ref{assump:misclassification} (i)--(ii) only concern the joint distribution of $T$ given $(T^*,z)$.
As such, they have the same meaning in a LATE model as in an additively separable model.
Imposing these conditions, Lemma \ref{lem:p_pstar} continues to hold in a LATE model.
It follows that $p_1 - p_0 = (1 - \alpha_0 - \alpha_1) (p_1^* - p_0^*)$ so that
\[
  \frac{\mathbb{E}[y|z=1] - \mathbb{E}[y|z=0]}{p_1 - p_0} = \frac{\mathbb{E}[Y(1) - Y(0)|J=c]}{1 - \alpha_0 - \alpha_1}.
\]
Moreover, $\alpha_0 \leq p_k \leq 1 - \alpha_1$ for all $k$.
Thus, the bound from Corollary \ref{cor:sharpBeta1} remains valid in a LATE model: $\mathbb{E}[Y(1) - Y(0)|J=c]$ must lie between the IV and reduced form estimands.

Unlike Assumptions \ref{assump:misclassification} (i)--(ii), Assumption \ref{assump:misclassification} (iii), non-differential measurement error, is explicitly stated in terms of the unobservable error term in an additively separable model.
Our derivation of the additional restrictions on $(\alpha_0, \alpha_1)$ implied by non-differential measurement error in the proof of Theorem \ref{thm:sharpII}, however, does not use Assumption \ref{assump:misclassification} (iii) directly.
Rather, it uses a condition that is \emph{equivalent} to it in an additively separable model, namely $\mathbb{E}[Y|T^*,T,z] = \mathbb{E}[Y|T^*,z]$.
Hence, as long as this equality holds, regardless of whether one is in an additively separable model or a LATE model, the bounds on $(\alpha_0, \alpha_1)$ from Theorem \ref{thm:sharpII} remain valid.
Since $Y = (1 - T^*)Y(0) + T^* Y(1)$, the appropriate modification of Assumption \ref{assump:misclassification} (iii) is as follows.
\begin{assump}[Non-differential Measurement Error]
  \label{assump:LATEnondiff}
  \[
    \mathbb{E}[Y(0)|T^*,T,z] = \mathbb{E}[Y(0)|T^*,z] \quad \mbox{and} \quad \mathbb{E}[Y(1)|T^*,T,z] = \mathbb{E}[Y(1)|T^*,z]
  \]
\end{assump}

To summarize, if one wishes to re-interpret our parameter $\beta$ as a local average treatment effect, the partial identification bounds from Theorems \ref{thm:sharpI} and \ref{thm:sharpII} above remain valid.
Assumption \ref{assump:model} (i) is replaced by $Y = h(T^*,z,\varepsilon)$, Assumption \ref{assump:model} (iii) is replaced by Assumptions \ref{assump:LATEtype}--\ref{assump:LATEmono}, and Assumption \ref{assump:misclassification} (iii) is replaced by Assumption \ref{assump:LATEnondiff}.
In a LATE model, however, our proofs of sharpness no longer apply, as they do not consider the testable implications of the LATE assumptions themselves.
For partial identification results that consider these implications but do not impose non-differential measurement error, see \cite{Ura}.
For discussion of the testable implications of a LATE model, see \cite{kitagawa}.

\small
\bibliographystyle{elsarticle-harv}
\bibliography{binary}
\normalsize

\end{document}